\documentclass[useAMS,usenatbib]{mn2e}
\usepackage{epsfig}
\usepackage{graphics}
\usepackage[usenames]{color}
\usepackage{amssymb}
\usepackage{amsmath}
\usepackage{times}
\usepackage{verbatim}
\usepackage{url}

\newcommand{\msun}{M_\odot}

\def\lsim{\lower.5ex\hbox{$\; \buildrel < \over \sim \;$}}

\def \plots {./plots/}

\title{Expected properties of the first gravitational wave signal detected with pulsar timing arrays}

\author[Rosado, Sesana \& Gair]{Pablo A. Rosado$^{1,2}$\thanks{E-mail: prosado@swin.edu.au}, Alberto Sesana$^{3,4}$, Jonathan Gair$^{5}$
\\
$^{1}$ Centre for Astrophysics \& Supercomputing, Swinburne University of Technology, PO Box 218, Hawthorn VIC 3122, Australia \\
$^{2}$ Max Planck Institute for Gravitational Physics, Albert Einstein Institute, Callinstra\ss e 38, 30167, Hanover, Germany \\
$^{3}$ Max Planck Institute for Gravitational Physics, Albert Einstein Institute, Am M\"uhlenberg 1, 14476, Golm, Germany \\
$^{4}$ School of Physics and Astronomy, The University of Birmingham, Edgbaston, Birmingham B15 2TT, UK \\
$^{5}$ Institute of Astronomy, Madingley Road, Cambridge, CB3 0HA, United Kingdom \\
}

\begin{document}

\date{}

\pagerange{\pageref{firstpage}--\pageref{lastpage}} \pubyear{2014}

\maketitle

\title{The nature of the first PTA detection}

\label{firstpage}

\begin{abstract}
In this paper we attempt to investigate the nature of the first gravitational wave (GW) signal to be detected by pulsar timing arrays (PTAs): will it be an individual, resolved supermassive black hole binary (SBHB), or a stochastic background made by the superposition of GWs produced by an ensemble of SBHBs?
To address this issue, we analyse a broad set of simulations of the cosmological population of SBHBs, that cover the entire parameter space allowed by current electromagnetic observations in an unbiased way.
For each simulation, we construct the expected GW signal and identify the loudest individual sources.
We then employ appropriate detection statistics to evaluate the relative probability of detecting each type of source as a function of time for a variety of PTAs; we consider the current International PTA, and speculate into the era of the Square Kilometre Array.
The main properties of the first detectable individual SBHBs are also investigated.
Contrary to previous work, we cast our results in terms of the detection probability (DP), since the commonly adopted criterion based on a signal-to-noise ratio threshold is statistic-dependent and may result in misleading conclusions for the statistics adopted here.
Our results confirm quantitatively that a stochastic signal is more likely to be detected first (with between 75 to 93 per cent probability, depending on the array), but the DP of single-sources is not negligible.
Our framework is very flexible and can be easily extended to more realistic arrays and to signal models including environmental coupling and SBHB eccentricity.
\end{abstract}


\section{Introduction}
Our current knowledge of the Universe is almost entirely based on electromagnetic observations (from radio waves, up to gamma rays).
A direct gravitational wave (GW) detection will open an alternative and complementary perspective on the Cosmos \citep{SathyaprakashSchutz2009}.
According to General Relativity, accelerating masses with a mass quadrupole moment varying in time generate GWs, and our Universe has plenty of systems fulfilling such a requirement \citep{Wald1983,Maggiore2008}; from compact binaries \citep[emitting GWs while inspiraling towards one another until coalescence,][]{PetersMathews1963}, to asymmetric spinning compact stars \citep{Broeck2005}, and catastrophic supernovae explosions \citep{YakuninEtAl2010}, just to mention a few.

A direct GW detection is a long-standing challenge in experimental physics, dating back to the '60s. 
To date, two different approaches to the problem have polarised the attention of the scientific community, bringing the promise of the first direct GW detection.
One technique is ground based laser interferometry; the second generation of ground based detectors (GBDs), including advanced LIGO \citep{LIGO2009}, advanced Virgo \citep{AccadiaEtAl2012} and KAGRA \citep{Somiya2012}, will soon be online, and is expected to observe tens of coalescing neutron star or stellar-mass black hole binaries \citep{AbadieEtAl2010}.
The other technique consists of timing an ensemble of ultra stable millisecond pulsars, forming what is commonly referred to as a pulsar timing array (PTA).
The times of arrival (ToAs) of radio pulses emitted by galactic milli-second pulsars are collected by 100-m class radio telescopes around the globe.
Since these pulses are so extremely regular, a GW passing by would introduce irregularities in their ToAs, a detectable fingerprint that can be measured, provided 100\,ns timing precision is achieved on a large number of pulsars \citep{Sazhin1978,Detweiler1979,HellingsDowns1983}.
Regardless of which approach, either GBDs or PTAs, achieves a GW detection first, the two independent detections are necessary, because they target orthogonal classes of sources, therefore providing complementary information about the Universe.
The invaluable scientific promises of GW astronomy are potentially so revolutionary that the design of a third generation of GBDs --the Einstein telescope (ET)-- is now being investigated \citep{PunturoEtAl2010}, and the European Space Agency (ESA) has selected \textit{The Gravitational Universe} science theme \citep{eLISA2013} for the L3 launch slot, with eLISA --the evolved laser interferometer space antenna-- put forward as strawman mission design, now scheduled for 2034.

Within the LIGO Scientific Collaboration\footnote{\url{http://www.ligo.org}}, different groups have formed in order to specialise in the detection of the four different types of expected GW signals: (i) a stochastic \textit{GW background} \citep[GWB,][]{LIGO2014}, (ii) \textit{continuous waves} \citep{LIGO2014b}, (iii) \textit{bursts} \citep{LIGO2014c}, and (iv) \textit{compact binary coalescences} \citep[aka CBC,][]{LIGO2013c}.
A GWB is the superposition of numerous similar GW sources, adding up in an incoherent fashion, and can have either an astrophysical origin, e.g. coming from a large population of a specific class of astrophysical objects \citep{Rosado2011,Regimbau2011}, or cosmological, e.g. coming from physical processes such as inflation or phase transitions in the early Universe \citep{Allen1996,Maggiore2000}.
Continuous waves are quasi-sinusoidal signals, expected to be produced by rotating neutron stars.
The burst search targets short GW transients ($\lesssim$ 1 second) with limited assumptions on the waveform.
On the contrary, the CBC search relies on accurate models for the waveform emitted by coalescing binaries, and seeks for patterns in the detector's data that match such models.

PTAs are following a similar path, but maintain a flexible structure.
There are three independent PTA consortia: the European PTA \citep[EPTA,][]{FerdmanEtAl2010}, the Parkes PTA \citep[PPTA,][]{ManchesterEtAl2013}, and NANOGrav \citep{JenetEtAl2009}.
The three groups join forces and combine data under the aegis of the International PTA \citep[IPTA,][]{HobbsEtAl2010}, which is formally a consortium of consortia.
The main target of PTA campaigns are the GWs emitted by supermassive black hole (SBH) binaries \citep[SBHBs; see, e.g.,][]{SesanaEtAl2008}, although processes in the early Universe may also produce GWs in the same frequency band, like cosmic strings \citep{SanidasEtAl2012} or the cosmological QCD phase transition \citep{CapriniEtAl2010}.

On the basis of the hierarchical paradigm of structure formation --according to which galaxies we see today underwent a series of merger events \citep{WhiteRees1978}-- and on the observational fact that SBHs are ubiquitous in galaxy centres \citep[e.g.,][]{MagorrianEtAl1998}, following galaxy mergers, a vast number of SBHBs is expected to form along the cosmic history \citep{BegelmanEtAl1980}.
Although there is plenty of observational evidence of the existence of SBH pairs separated by hundreds-to-thousands of parsecs, detecting parsec and sub-parsec scale SBHBs driven by GWs has been proven challenging, and only candidate systems supported by circumstantial evidence have been proposed to date \citep[see][for an extensive review on the topic]{DottiEtAl2012}. The recent access to vast time domain optical data set resulted in the discovery of a number of quasars showing periodic variability \citep{GrahamEtAl2015,LiuEtAl2015}, but the SBHB nature of these objects is difficult to prove.
In any case, a large population of SBHBs is an inevitable prediction of current galaxy formation models, and their GW signals must have travelled to our galaxy imprinting their signature in the pulsars' ToAs.
The incoherent superposition of many signals will possibly result in a stochastic GWB \citep{RajagopalRomani1995}, but particularly massive and/or nearby SBHBs might dominate the signal budget and show up as resolvable continuous waves \citep{SesanaEtAl2009} in the data.
Moreover, bursts from particularly eccentric binaries \citep{AmaroSeoaneEtAl2010,FinnLommen2010} or from the GW memory effect \citep{Favata2009,PshirkovEtAl2010,VanHaasterenLevin2010}  are also possible.
Consequently, several IPTA projects have been established, targeting different types of signals.
Unlike GBDs, there are no CBC searches within the PTA, since SBHBs are observed long before coalescence, during their slow, adiabatic inspiral.
For PTAs, the first detection will probably be either a continuous wave signal from a SBHB (we will refer to such signal as a 'single source' from now on) or a GWB, and we focus on these two classes of sources in this study.

A GWB is commonly characterised by an amplitude and a spectral slope \citep{JenetEtAl2005}.
Detecting a GWB, with amplitude and slope consistent with the expected signal generated by an ensemble of SBHBs, would prove the validity of the models that predict the existence of tight SBHBs, and, in a broader context, would support our current understanding of the formation and evolution of galaxies.
Moreover, the detection of a GWB could put constraints on alternative theories of gravity, or prove their validity against General Relativity \citep{ChamberlinSiemens2012}.
However, an ensemble of SBHBs could also produce a spectrum different than a simple power law \citep{SesanaEtAl2004,KocsisSesana2011,Sesana2013b,McWilliamsEtAl2014,RaviEtAl2014}, and other GW sources could also lead to different measurable GWBs; detecting the actual shape of the GWB would help discriminate between these different models \citep{SampsonEtAl2015}.

The detection of a single SBHB \citep{YardleyEtAl2010,CorbinCornish2010,Ellis2013,WangEtAl2014} would allow for some further appealing prospects.
First, by detecting a purely stochastic GWB one cannot, in principle, be sure of the kind of sources that originated it; however, detecting a single SBHB would be a more direct proof of the existence of tight binaries.
Detecting a single source would allow us to measure the characteristics of the binary, for example its luminosity distance, mass, orbital period, and sky location, opening fascinating prospects for multimessenger astronomy \citep{SesanaEtAl2012,TanakaEtAl2012,BurkeSpolaor2013,RosadoSesana2014}.
In some cases, it could also help to improve our measurement of the distance to the pulsars in the array \citep{LeeEtAl2011}.

Although it is commonly believed that a GWB detection will likely precede the identification of any single source, this statement has never been properly quantified in the literature.
The aim of this paper is precisely to answer the question: what type of signal will more likely be detected first by a PTA, and with what probability?
To tackle the problem, we analyse a large set of simulated realisations of the ensemble of SBHBs, and compare the detectability of the two types of signals as a function of time in each of them.
The realisations are produced following the prescriptions of \cite{Sesana2013}, employing, for this exploratory study, the simplifying assumption of circular, GW-driven binaries.

The outline of the paper is as follows.
In Section \ref{sc:models} we describe the simulations of the SBHB cosmic population and present the necessary mathematical tools to evaluate the detectability of the GWB and of single SBHBs; finally, we describe the properties of the pulsar arrays assumed as GW detectors.
In Section \ref{sc:results}, the detection probability of the GWB is compared to that of single SBHBs, assuming the IPTA and two configurations of the Square Kilometre Array (SKA); we also study the signal-to-noise ratio (S/N) of the two types of signals, and comment on the implications of evaluating their detectability using the S/N.
We then investigate the properties of the first detectable single binaries, including the shift between Earth and pulsar terms of the signal.
In Section \ref{sc:discussion} we comment on some of the main assumptions adopted and caveats of the paper, and summarise our findings.
Additionally, in Appendix \ref{app:stats} we include a derivation of the formulas used to evaluate the detection of the GWB.

\section{GW signal models and detection statistics}
\label{sc:models}

\subsection{Observationally-based GW simulations}
\label{sec:sims}

To construct the relevant SBHB population emitting in the PTA frequency band, we exploit the observationally-based approach put forward by \cite{Sesana2013}.
We assume for simplicity that SBHBs are all circular and GW-driven \citep[see][for a discussion of the impact of binary-environment coupling and eccentricity]{Sesana2013b}.

In practice, the SBHB population depends on four ingredients:
\begin{enumerate}
\item {\it The galaxy merger rate}.
The differential merger rate can be written as
\begin{equation}
\frac{d^3n_G}{dzdMdq}=\frac{\phi(M,z)}{M\ln{10}}\frac{{\cal F}(z,M,q)}{\tau(z,M,q)}\frac{dt_r}{dz}.
\label{galmrate}
\end{equation}
Here, $\phi(M,z)=[dn/d{\rm log}M]_z$ is the galaxy mass function measured at redshift $z$; ${\cal F}(M,q,z)=[df/dq]_{M,z}$ is the differential fraction of galaxies with mass $M$ at redshift $z$ paired with a secondary galaxy having a mass ratio in the range $q, q+\delta{q}$; $\tau(z,M,q)$ is the typical merger time-scale for a galaxy pair with a given $M$ and $q$ at a given $z$; and $dt_r/dz$ converts the proper time rate into redshift, and is given by standard cosmology.
The reason for writing Equation (\ref{galmrate}) is that $\phi$ and ${\cal F}$ can be directly measured from observations, whereas $\tau$ can be inferred by detailed numerical simulations of galaxy mergers.
We take five different galaxy stellar mass functions from the literature \citep{BorchEtAl2006,DroryEtAl2009,IllbertEtAl2010,MuzzinEtAl2013,TomczakEtAl2014} and match them with the local mass function \citep{BellEtAl2003}, to obtain five fiducial $\phi_z(M)$.
We consider four studies of the evolution of the galaxy pair fraction \citep{BundyEtAl2009,DeRavelEtAl2009,LopezSanjuanEtAl2012,XuEtAl2012}.
Finally, we take merger time-scales ${\tau}$ estimated by \cite{KitzbichlerWhite2008} and \cite{LotzEtAl2010}.
\item {\it The relation between SBHs and their hosts}.
We assign to each merging galaxy pair two SBHs with masses drawn from 13 different SBH-galaxy relations found in the literature \citep{HaringRix2004,GultekinEtAl2009,SaniEtAl2011,GrahamEtAl2011,Graham2012,GrahamScott2013,BeifioriEtAl2012,McConnellMa2013,KormendyHo2013}, spanning a broad range of uncertainty and including recent observations that corrected SBH estimates upwards.
\item {\it The efficiency of SBH coalescence following galaxy mergers}.
We simply assume that SBHBs coalesce efficiently following galaxy mergers, therefore bypassing the `last parsec problem' \citep{MilosavljevicMerritt2003}.
As already stated, we do not dig into complications related to the SBHB-environment coupling \citep{KocsisSesana2011,Sesana2013b,McWilliamsEtAl2014,RaviEtAl2014}, and assume that all systems are circular and GW-driven.
\item {\it When and how accretion is triggered during a merger event}.
We allow, during the merger, for some amount of accretion on each SBH.
This mass can be accreted with a different timing with respect to the SBH binary merger, and in different amount on the two SBHs, following the scheme described in Section 2.2 of \cite{SesanaEtAl2009}.
\end{enumerate}

We combine the different ways to populate the merging galaxies with SBHs together with the galaxy merger rates to obtain 23400 different SBHB merger rates, consistent with current observations of the evolution of the galaxy mass function and pair fractions at $z<1.3$ and $M>10^{10}\msun$ and with the empirical SBH-host relations published in the literature.
We give equal credit to each model, and run 10 realisations of each, producing a total of 234000 simulated universes, each one with a particular GW signal produced by the ensemble of SBHBs. 
This number of realisations is sufficient to place reasonable confidence levels for the expected amplitude according to {\it current observational constraints}.
Our approach is modular in nature, and it is straightforward to expand the range of models to include new estimates of all the quantities involved.

The output of each realisation of the Universe is a list of SBHBs including for each system:
\begin{enumerate}
\item $\mathcal{M}$, the proper chirp mass, which is related to the individual SBH masses $m_1$ and $m_2$ by $\mathcal{M}=[m_1 m_2]^{3/5}/[m_1+m_2]^{1/5}$,
\item $f$, the observed (redshifted) GW frequency,
\item $z$, the observed redshift.
\end{enumerate}
From the redshift, the comoving distance to a binary is calculated by assuming a $\Lambda$ cold dark matter universe,
\begin{equation}
r=\frac{c}{H_{100} h}\int_0^z \left[ \Omega_m[1+z]^3+\Omega_\Lambda \right]^{-1/2} dz,
\end{equation}
where $c$ is the speed of light, $H_{100}=100\,$km\,s$^{-1}$\,Mpc$^{-1}$, and the cosmological parameters adopted are
\begin{equation}
(h,\Omega_m,\Omega_\Lambda)=(0.7,0.3,0.7).
\end{equation}
To determine the strength of the GW signal detected on Earth, other parameters need to be specified, namely:
\begin{itemize}
\item $(\phi, \theta)$, longitudinal and latitudinal spherical coordinates describing the SBHB location in the sky, 
\item $\iota$, the binary inclination angle with respect to the line of sight,  
\item $\psi$, the GW polarisation angle, 
\item $\Phi_0$, the initial GW phase. 
\end{itemize}
When simulating the signal from single sources, the previous quantities will be obtained by drawing numbers from the appropriate uniform distributions.
On the other hand, when simulating the GWB signal, we will simply use the sky and polarisation average strain \citep{2000PhRvD..62l4021F},
\begin{equation}
\label{eq:strain}
h=A\sqrt{\frac{1}{2}\left[a^2+b^2\right]},
\end{equation}
where 
\begin{equation}
\label{eq:ampli}
A=2\frac{G^{5/3}\mathcal{M}^{5/3}[\pi f [1+z]]^{2/3}}{c^4 r}
\end{equation}
is the dimensionless amplitude of the signal, and the contributions from the two wave polarisations are defined by
\begin{equation}
a=1+\cos^2{\iota},
\end{equation}
and
\begin{equation}
b=-2\cos{\iota}.
\end{equation}
One can note that, using the previous formulas, the strain of the GWB is independent of the binaries' exact sky location and polarisation angle.
As discussed in the next section, this is likely to have only a minor impact on our results.

\begin{figure}
\includegraphics{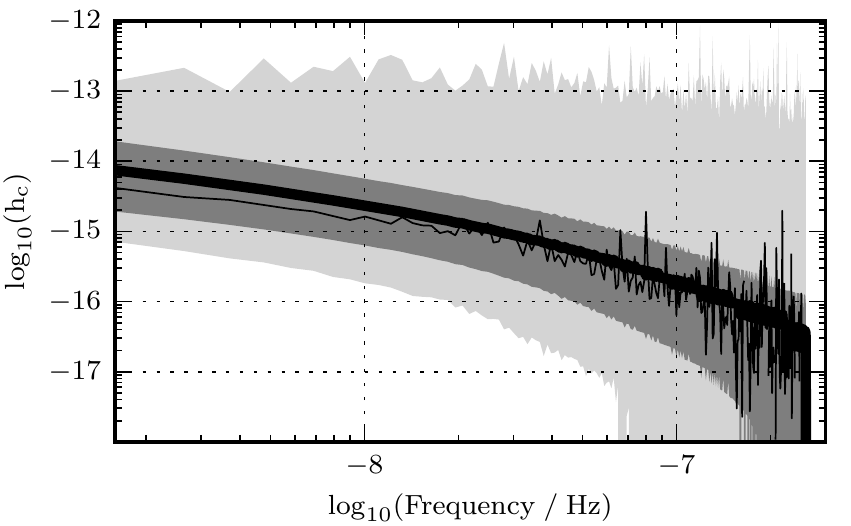}
\caption{Characteristic GW strain versus observed GW frequency.
For each realisation of the Universe we obtain a curve $h_c(f)$, as the sum of the contribution from all binaries (the thin black line shows the output of one particular realisation).
The light-grey area contains all possible values of $h_c$ found in the realisations, whereas the dark-grey area brackets the 5th and 95th percentiles of the $h_c$ distribution.
The thick black line is the mean $h_c$ at each frequency over all realisations.
A frequency bin of size $\Delta f=[30\,\text{yr}]^{-1}$ has been assumed.}
\label{fg:hc}
\end{figure}

If we time an ensemble of millisecond pulsars for a period $T$, the overall amplitude of an incoherent superposition of GWs can be described in terms of a characteristic GW strain $h_c$ at each observed frequency, which is related to the strain of the individual sources via
\begin{align}
\label{eq:hc}
h_c^2=\frac{\sum_k h_k^2 f_k}{\Delta f},
\end{align}
where $k$ is an index running over all sources in a given frequency bin of width $\Delta{f}=1/T$.
An overview of the simulated signals is presented in Figure \ref{fg:hc}, in which we show $h_c$ for all the realisations of the SBHB cosmic population.
The light grey area spans the range of all possible $h_c$ found in the realisations, whereas the 5th and 95th percentiles of the $h_c$ distribution are contained in the dark grey area.
One can notice that $h_c$ reaches values above $10^{-13}$ at almost all frequency bins. This is because, over more than $2\times10^5$ realisations of the Universe, it is likely to find some extremely massive and nearby SBHB.
However, such high strain values can be regarded as rare outliers, whereas the region between the 5th and the 95th percentile of the distribution is much narrower.
The black thick line gives the mean $h_c$ over all realisations at each frequency bin, which is consistent with previous predictions for the amplitude of the GWB \citep{RajagopalRomani1995,WyitheLoeb2003,SesanaEtAl2008,Rosado2011,RaviEtAl2015}.
One example of an individual realisation is also added in the figure, plotted with a thin black line.
The size of the frequency bin is $\Delta f=[30\,\text{yr}]^{-1}$, which corresponds to the longest observing time we will consider in this work.
At each individual frequency, $h_c$ can be either dominated by a single loud source, or produced by a superposition of several SBHBs, each contributing a sizeable share of the GW strain.
Consequently, for detection purposes, the signal might be either deterministic or incoherent/stochastic in nature.
We now turn to the description of the detection strategies adopted in the two cases, which is necessary to assess which kind of signal will likely be detected first. 

\subsection{Detection of a stochastic background}
\label{sc:gwb}

Let us assume that we have a large set of realisations of the Universe, all of them with similar astrophysical properties\footnote{Throughout this section, by `realisations' we do not refer to the outputs of the computer simulations analysed in other sections of this paper, but to a hypothetical set of copies of the same Universe.}.
When searching for a GWB, we define our detection statistic $S$ as the cross-correlation between the outputs of two detectors (two pulsars\footnote{We assume that the optimal way to cross-correlate many detectors is to combine them in pairs \citep{AllenRomano1999}.}).
After a certain observing time, each realisation of the Universe has a measurement of $S$.
We assume that the collection of values of $S$ from different realisations is a stochastic process.

In the absence of a GWB, the cross-correlation output reflects the properties of the noise processes involved in the measurement.
We assume this to be a stochastic Gaussian process with probability density function (PDF) defined by a mean $\mu_0$ and a standard deviation $\sigma_0$,
\begin{equation}
p_0(S)=\frac{1}{\sqrt{2\pi\sigma_0^2}}e^{-\frac{[S-\mu_0]^2}{2\sigma_0^2}}.
\end{equation}
We further assume that the noise in all detectors is white, stationary, with zero mean ($\mu_0=0$), and uncorrelated.
If, on the other hand, a GWB is present in the data (the same GWB in all realisations), the detection statistic will follow a different distribution, namely
\begin{equation}
p_1(S)=\frac{1}{\sqrt{2\pi\sigma_1^2}}e^{-\frac{[S-\mu_1]^2}{2\sigma_1^2}},
\end{equation}
where the mean $\mu_1$ is now larger than zero, and the standard deviation $\sigma_1$ is, in general, different than $\sigma_0$.

Given a certain value of $S$ measured in one realisation, we claim that it may contain a GWB if $S\ge S_\text{T}$, where $S_\text{T}$ is a pre-defined detection threshold.
The integral of $p_0(S)$ over all values of $S\ge S_\text{T}$ gives the \textit{false alarm probability} (FAP),
\begin{equation}
\label{eq:alphaint}
\alpha = \int_{S_\text{T}}^{\infty} p_0(S) dS,
\end{equation}
which is the probability of claiming a spurious detection in the absence of a GWB.
Alternatively, the integral of $p_1(S)$ over all values of $S\ge S_\text{T}$ gives the \textit{detection probability} (DP),
\begin{equation}
\label{eq:gammaint}
\gamma = \int_{S_\text{T}}^{\infty} p_1(S) dS,
\end{equation}
which is the probability of claiming a true detection of the GWB when it is indeed present.

Introducing the complementary error function (erfc), we can solve the integrals of Equations (\ref{eq:alphaint}) and (\ref{eq:gammaint}), to obtain
\begin{equation}
\label{eq:alpha}
\alpha=\frac{1}{2}\text{erfc}\left( \frac{S_\text{T}}{\sqrt{2}\sigma_0}\right),
\end{equation}
and
\begin{equation}
\label{eq:gammabg}
\gamma=\frac{1}{2}\text{erfc}\left( \frac{S_\text{T}-\mu_1}{\sqrt{2}\sigma_1}\right).
\end{equation}
Throughout the paper we fix the FAP to the value $\alpha_0=0.001$.
We can then solve for $S_\text{T}$ in Equation (\ref{eq:alpha}) and replace the result into Equation (\ref{eq:gammabg}) to get
\begin{equation}
\label{eq:DP}
\gamma_\text{B}=\frac{1}{2}\text{erfc}\left[ \frac{\sqrt{2}\sigma_0\text{erfc}^{-1}(2\alpha_0)-\mu_1}{\sqrt{2}\sigma_1}\right].
\end{equation}
This is the quantity that we will use to evaluate the detectability of a GWB. Later on we will explicitly give formulas for $\mu_1$, $\sigma_0$, and $\sigma_1$, but first there is an important aspect of the statistic $\mathcal{S}$ that needs to be commented on.

The cross-correlation $S$ is constructed as follows,
\begin{equation}
S=\int_{-T/2}^{T/2}dt \int_{-T/2}^{T/2} dt' s_i(t) s_j(t')Q(t,t'),
\label{eq:crosscorr}
\end{equation}
where $T$ is the observation time, $s_i(t)$ and $s_j(t')$ are the data from two different pulsars, and $Q(t,t')$ is a filter function.
The latter must be chosen in such a way that the DP is maximised for a fixed value of FAP; in other words, we adopt the Neyman-Pearson criterion to define our statistics.
Assuming that $S$ follows a Gaussian distribution, one can obtain closed formulas for $\mu_1$, $\sigma_0$ and $\sigma_1$ as a function of $Q(t,t')$, replace them in Equation (\ref{eq:DP}), and obtain the filter function that maximises $\gamma_\text{B}$.
However, the maximisation of the DP is a non-trivial task.
Instead, it is customary to construct statistics that are optimal in the sense of maximising a proxy for the DP, which is the S/N.
The latter can be defined in two different ways, $\mu_1/\sigma_0$, or $\mu_1/\sigma_1$, with different implications.
Hence, there are two possible ways to construct the optimal statistic:
\begin{itemize}
\item Maximise $\text{S/N}_\text{A}=\mu_1/\sigma_0$. We will refer to the statistic constructed in this way as the \textit{A-statistic}.
\item Maximise $\text{S/N}_\text{B}=\mu_1/\sigma_1$. We will refer to this as the \textit{B-statistic}.
\end{itemize}
The derivation of these two statistics and their properties are discussed in Appendix~\ref{app:stats}.

If the signature of the GWB in the data is small, i.e. in the \textit{weak signal approximation}, $\sigma_0$ and $\sigma_1$ are almost identical, and the two statistics become equivalent.
This assumption is usually adopted in the literature, especially for GBD studies \citep{AllenRomano1999,Rosado2012,RegimbauEtAl2014}, and allows us to define the signal-to-noise ratio as S/N$=\mu_1/\sigma_0\approx \mu_1/\sigma_1$.
Under this assumption, the true optimal filter for the cross-correlation is in fact the one that maximises the S/N.
Furthermore, one can set $\sigma_0\approx \sigma_1$ in Equation (\ref{eq:DP}) and fix the DP to a particular value $\gamma_0$ to obtain
\begin{equation}
\text{S/N}^\text{T}\approx \sqrt{2}\left[ \text{erfc}^{-1} (2\alpha_0) - \text{erfc}^{-1}(2\gamma_0) \right].
\end{equation}
If for example we set $\alpha_0=0.001$ and $\gamma_0=0.95$, S/N$^\text{T}\approx 4.74$ is the threshold such that a larger value of S/N ensures a FAP smaller than $0.1\%$ and a DP larger than $95\%$, which could be considered a confident detection.

However, \cite{SiemensEtAl2013} already pointed out that the weak signal approximation is, in the long run, bound to become inaccurate for PTAs; and since we want to make predictions for future, more sensitive arrays, the approximation would certainly be crude.
If the presence of a signal is tangible, and therefore $\sigma_0$ and $\sigma_1$ are not equal (the latter being typically an increasing function of time), one cannot associate pre-defined values of $\alpha_0$ and $\gamma_0$ to a fixed S/N threshold.
Instead, one has two possible ways to define the S/N threshold:
\begin{itemize}
\item If one adopts the A-statistic, S/N$_\text{A}=\mu_1/\sigma_0$.
The threshold would then be
\begin{equation}
\text{S/N}^\text{T}_\text{A}= \sqrt{2}\left[ \text{erfc}^{-1} (2\alpha_0) - \frac{\sigma_1}{\sigma_0}\text{erfc}^{-1}(2\gamma_0) \right].
\end{equation}
\item If one instead adopts the B-statistic, S/N$_\text{B}=\mu_1/\sigma_1$.
The threshold would then be
\begin{equation}
\label{eq:snrbstat}
\text{S/N}^\text{T}_\text{B}= \sqrt{2}\left[ \frac{\sigma_0}{\sigma_1} \text{erfc}^{-1} (2\alpha_0) - \text{erfc}^{-1}(2\gamma_0) \right].
\end{equation}
\end{itemize}
In both cases, as one can see, the threshold changes in time as $\mu_1$ and $\sigma_1$ evolve. More specifically, for fixed DP and FAP, the S/N threshold increases for the A-statistic and decreases for the B-statistic.

To avoid confusion, we prefer to evaluate the detectability of the signal in terms of the DP directly, avoiding the necessity of defining an S/N and a certain threshold.
In the following we adopt the B-statistic, for reasons discussed in Appendix~\ref{app:stats}.
This is given by
\begin{align}
&X_\text{B}= \nonumber \\
&\sum_k \sum_{ij} \frac{2 \Gamma_{ij} S_{h0}(f_k)  s_{ijk}} {[P_i(f_k) +S_{h0}(f_k)][P_j(f_k)+S_{h0}(f_k)] +\Gamma_{ij}^2 S_{h0}^2(f_k)},
\label{eq:statBdef}
\end{align}
which is a linear combination of the cross-correlations between the data from pulsars $i$ and $j$ at discrete frequency $f_k$,
\begin{align}
s_{ijk}=\tilde{s}_i^*(f_k) \tilde{s}_j(f_k).
\end{align}
Here, the `\ $^*$\ ' and `\ $\tilde{}$\ ' denote the complex conjugate and Fourier transform, respectively. 
The detailed derivation of Equation (\ref{eq:statBdef}) along with the formal definition of the cross-correlation is given in Appendix ~\ref{app:stats}, where we also derive the three quantities that are necessary in order to evaluate the DP in Equation (\ref{eq:DP}), namely $\mu_1$, $\sigma_0$, and $\sigma_1$. In the following we will define the various quantities introduced in Equation (\ref{eq:statBdef}).

The summation in $k$ is over all frequency bins between $f_\text{min}=T^{-1}$, and $f_\text{max}=[\Delta t]^{-1}$, where the size of each bin is $\Delta f=T^{-1}$, and the frequency $f_k$ is assumed at the arithmetic mean of bin $k$.
The parameter $\Delta t$ is the cadence time, i.e. the typical time lapsed between consecutive pulsar observations.
We assume a cadence time of 2 weeks for all pulsars and arrays.
For such a cadence, one has $f_\text{max}\approx 2\times 10^{-6}\,$Hz, but it is extremely unlikely to find a SBHB emitting at such a high GW frequency, and we use $f_\text{max}\approx 3\times10^{-7}$\,Hz instead in our computation.
The observation time $T$ is the duration of a PTA campaign; for example, the current IPTA has been recording ToAs of radio pulses for $\sim$10\,yr.
The other summation in the Equation (\ref{eq:statBdef}) accounts for all possible pulsar pairs,
\begin{equation}
\sum_{ij}=\sum_{i=1}^M \sum_{j>i}^M,
\end{equation}
where $M$ is the number of pulsars in the array, which is, for the current IPTA, $M=49$; when considering SKA-type arrays we will assume different values for $M$.
$P_i$ is the noise power spectrum of the $i$-th pulsar; we assume that the pulses have a certain degree of irregularity, which is a random Gaussian process described by a root mean square (rms) value $\sigma_i^2$, so that $P_i$ is simply
\begin{align}
\label{eq:pisimple}
P_i=2\sigma_i^2\Delta t.
\end{align}
Typical values of $\sigma$ are between 500\,ns and 5\,$\mu$s (which correspond to the best and worst case within the IPTA, respectively).
$\Gamma_{ij}$ is the overlap reduction function \citep{FinnEtAl2009,ThraneRomano2013}, that for a PTA has the form \citep{HellingsDowns1983}
\begin{align}
\label{eq:gamma}
&\Gamma_{ij}=\frac{3}{2}\gamma_{ij}\ln \left(\gamma_{ij}\right)-\frac{1}{4}\gamma_{ij}+\frac{1}{2}+\frac{1}{2}\delta_{ij},
\end{align}
where $\gamma_{ij}=[1-\cos(\theta_{ij})]/2$, and $\theta_{ij}$ is the relative angle between pulsars $i$ and $j$.
The term multiplying the Kronecker Delta $\delta_{ij}$ is irrelevant in the calculations, and is introduced just to normalise the overlap reduction function in such a way that $\Gamma_{ii}=1$.
A full mathematical derivation of $\Gamma_{ij}$ can be found in \cite{AnholmEtAl2009}.
$S_h$ is the one-sided power spectral density of the GW signal in the timing residuals,
\begin{equation}
\label{eq:sh}
S_h=\frac{h_c^2}{12\pi^2f^3},
\end{equation}
where $h_c$ is the characteristic GW strain given by Equation (\ref{eq:hc}).
Finally, $S_{h0}$ is the expected one-sided power spectral density of the GW signal, which is needed to construct the optimal statistic.
A simplistic assumption regarding the signal would be that it follows a power law of known slope,
\begin{equation}
\label{eq:sh0}
S_{h0}=\frac{\mathcal{A}^2 \text{yr}^{-4/3}}{12\pi^2} f^{-13/3},
\end{equation}
where $\mathcal{A}$ would be a fiducial characteristic strain amplitude, e.g. $10^{-15}$.
With this definition, the corresponding characteristic strain of the GWB would be $h_c=\mathcal{A} [f/\text{yr}^{-1}]^{-2/3}$, which has the expected average frequency dependence of Equation (\ref{eq:hc}) in the limit of a smooth signal.
However, we found that in practice there is little difference in the performance of the statistic between using the fiducial model of Equation (\ref{eq:sh0}) and just setting $S_{h0}=S_h$ for each realisation of $S_h$.
Hence, throughout the rest of the paper we assume that $S_{h0}$ is identical to $S_h$, given in Equation (\ref{eq:sh}).
This choice can be intricate in practice, when applied to a real detection pipeline (given that some prior knowledge of the shape of the yet undetected signal is required), but is convenient for the purposes of this work.
In Appendix \ref{app:stats} we justify this choice and comment on its practical implications.

By working purely with the power in each frequency bin, $S_h$, in our treatment of the stochastic background we are effectively smearing out each source so that the emission is isotropic over the sky.
In reality, each signal is a point source and so the GW power distribution on the sky will be anisotropic.
Using an isotropic search when the background is anisotropic is sub-optimal and will have a lower DP.
Techniques for searching for and characterising anisotropic backgrounds have been developed~\citep{GairEtAl2014}, which could out-perform isotropic searches in certain regimes, although this has not yet been investigated fully.
The DP of an optimal anisotropic background search is likely to be lower than an optimal isotropic search for an isotropic background of the same net amplitude, due to the larger number of model parameters in the anisotropic case.
For these reasons our results may be slightly overestimating the detectability of the stochastic background.
However, the background only deviates significantly from isotropy in frequency bins that contain only a small number of sources, which are mostly the higher frequency bins.
The signal to noise ratio for the isotropic background search tends to be dominated by the lower frequency bins and so it is likely that any overestimate of the DP is fairly small.
Nonetheless, this should be investigated further in the future, using simulations in which the contribution from each individual binary is separately added into the residuals for each pulsar.

\subsection{Detection of a single source}
\label{sc:single}

The optimal way to search for a deterministic signal whose parameters are unknown is to adopt a \textit{matched filter} \citep{Schutz1997}.
The waveform of a circular binary is generally described by $7+2M$ parameters.
The Earth term (a single sinusoidal GW) is fully described by 7 parameters: ${(h,f,\theta,\phi,\psi,\iota,\Phi_0)}$, already introduced in Section \ref{sec:sims}, and each of the $M$ pulsar terms adds an additional GW frequency and phase ${(f_i,\Phi_i)}$ to the list of parameters.
The construction of an adequate detection statistic depends on the functional form of the template, and is different for evolving and non-evolving binaries (i.e., binaries for which $f_i<f$ or $f_i=f$, respectively).
For the sake of simplicity, we assume binaries with orbital evolution time-scales shorter than the typical pulsar-Earth light travel time (i.e. evolving binaries). For such systems, the pulsar and Earth terms fall at different frequencies. One can then construct a simple template for the coherent combination of the Earth terms only, discarding possible contribution to the signal from the pulsar terms.
In this case, it has been shown that the relevant parameter space for the construction of the signal template can be reduced to three dimensions only; namely the frequency $f$ and the sky location $\theta,\phi$.
The resulting $\mathcal{F}_e$-statistic is optimal in the Neyman-Pearson sense \citep{Jaranowski2005,BabakSesana2012,Ellis2012}.
In the absence of a signal, the PDF of the $\mathcal{F}_e$-statistic follows a $\chi^2$ distribution with 4 degrees of freedom,
\begin{equation}
p_0(\mathcal{F}_e)=\mathcal{F}_e e^{-\mathcal{F}_e},
\end{equation}
and if the signal is present, the PDF is a non-central $\chi^2$ distribution with 4 degrees of freedom,
\begin{equation}
\label{eq:p1}
p_1(\mathcal{F}_e,\rho)=\frac{[2\mathcal{F}_e]^{1/2}}{\rho}I_1(\rho\sqrt{2\mathcal{F}_e})e^{-\mathcal{F}_e-\frac{1}{2}\rho^2}.
\end{equation}
The function $I_1(x)$ is the modified Bessel function of the first kind of order 1, and the non-centrality parameter $\rho$ is exactly equal to the optimal signal-to-noise ratio S/N$_\text{S}$ (whose calculation will be explained later in this section).

Let us assume for a moment that we know the intrinsic parameters of the signal we are searching for, namely $f$, $\theta$, and $\phi$.
In this case, the FAP can be simply calculated by integrating the PDF of the statistic in the absence of signal as
\begin{equation}
\label{eq:alphai}
\alpha_i=\int_{\bar{\mathcal{F}_e}}^{\infty}p_0(\mathcal{F}_e)d\mathcal{F}_e=[1+\bar{\mathcal{F}_e}]e^{-\bar{\mathcal{F}_e}}.
\end{equation}
However, if those intrinsic parameters are unknown (which is the most general case), one has to filter the data with a number of templates $N$ that cover the relevant parameter space of possible GW signals.
Each template is an independent trial, and since we are now performing the same experiment $N$ times, the total FAP becomes
\begin{equation}
\label{eq:alphas}
\alpha=1-[1-\alpha_i]^{N},
\end{equation}
where the index $i$ identifies the FAP in the single trial case.
The total FAP is therefore function of the exact choice of the number of templates $N$, which will be discussed later.
As in the previous section, we can obtain the threshold in the statistic by choosing a certain value of FAP, $\alpha_0$.
Then, introducing Equation (\ref{eq:alphai}) in (\ref{eq:alphas}),
\begin{equation}
\alpha_0=1-[1-[1+\bar{\mathcal{F}_e}]e^{-\bar{\mathcal{F}_e}}]^{N},
\end{equation}
which allows us to numerically obtain the threshold $\bar{\mathcal{F}_e}$.
On the other hand, the DP can be calculated by numerically integrating Equation (\ref{eq:p1}),
\begin{align}
\label{eq:gammas}
\gamma_i&=\int_{\bar{\mathcal{F}_e}}^{\infty}p_1(\mathcal{F}_e,\rho)d\mathcal{F}_e \nonumber\\
&=\int_{\bar{\mathcal{F}_e}}^\infty \frac{[2\mathcal{F}_e]^{1/2}}{\rho}I_1(\rho\sqrt{2\mathcal{F}_e})e^{-\mathcal{F}_e-\frac{1}{2}\rho^2} d\mathcal{F}_e.
\end{align}
This is the probability of detecting one binary (in a particular frequency bin).
But we are not interested in any specific binary, and, as a matter of fact, we should consider any potentially resolvable SBHB in our DP calculation.
So, if $\gamma_i$ is the DP of a single source in a specific frequency bin, then the total probability of detecting at least one single source in any frequency bin is given by
\begin{equation}
\label{eq:gammastotal}
\gamma_\text{S}=1-\prod_i [1-\gamma_i],
\end{equation}
where the index $i$ includes all frequency bins between $f_\text{min}$ and $f_\text{max}$.
Equation (\ref{eq:gammastotal}) gives the DP of a single source, in analogy to Equation (\ref{eq:DP}), that was the DP of a GWB; these two equations are the main quantities that we need to investigate in order to compare the detectability of a GWB and single sources.
In order to solve the integral in Equation (\ref{eq:gammas}), we first need to know how to calculate $\rho=$S/N$_\text{S}$, which we now explain.

In each realisation of the ensemble of SBHBs, we select the \textit{strongest} binary in each frequency bin.
By this we mean the binary that produces the largest single characteristic strain within the bin,
\begin{align}
\label{eq:hcstrongest}
h_c^\text{max}=\max\left( \sqrt{\frac{h_b^2 f_b}{\Delta f}} \right),
\end{align}
where the index $b$ runs over all the binaries within the same frequency bin in a particular realisation.
The optimal S/N for the Earth term signal of a SBHB, in an array of $M$ pulsars, is given by the coherent addition of the signal in each individual pulsar \citep{SesanaVecchio2010}
\begin{equation}
\label{eq:snrs}
\text{S/N}_\text{S}=\left[ \sum_{i=1}^{M} \text{S/N}_i^2 \right]^{1/2},
\end{equation}
where 
\begin{align}
\label{eq:snralpha}
&\text{S/N}_i^2=\frac{2}{S_i} \frac{A^2}{4\pi^2f^2} \nonumber\\
&\times \int_0^T \left[ a F_i^+ [\sin(\Phi)-\sin(\Phi_0)]-b F_i^\times [\cos(\Phi)-\cos(\Phi_0)]\right]^2 dt.
\end{align}
Here there are several quantities to be defined.
The amplitude $A$ is the one introduced in Equation (\ref{eq:ampli}).
The GW phase is
\begin{equation}
\Phi=2\pi f t.
\end{equation}
The antenna pattern functions are
\begin{equation}
F_i^+=\frac{1}{2}\frac{[\hat{m}\cdot \hat{p}_i]^2-[\hat{n}\cdot \hat{p}_i]^2}{1+\hat{\Omega}\cdot \hat{p}_i},
\end{equation}
and
\begin{equation}
F_i^\times=\frac{[\hat{m}\cdot \hat{p}_i][\hat{n}\cdot \hat{p}_i]}{1+\hat{\Omega}\cdot \hat{p}_i},
\end{equation}
where the unitary vectors introduced are
\begin{align}
\hat{m}=&+[\sin(\phi)\cos(\xi)-\sin(\xi)\cos(\phi)\cos(\theta)] \hat{x} \nonumber\\
& - [\cos(\phi)\cos(\xi)+\sin(\xi)\sin(\phi)\cos(\theta)] \hat{y} \nonumber\\
& + [\sin(\xi)\sin(\theta)] \hat{z},
\end{align}
\begin{align}
\hat{n}=&+[-\sin(\phi)\sin(\xi)-\cos(\xi)\cos(\phi)\cos(\theta)] \hat{x} \nonumber\\
&+ [\cos(\phi)\sin(\xi)-\cos(\xi)\sin(\phi)\cos(\theta)] \hat{y} \nonumber\\
&+ [\cos(\xi)\sin(\theta)] \hat{z},
\end{align}
\begin{align}
\hat{\Omega}=-\sin(\theta)\cos(\phi) \hat{x}
- \sin(\theta)\sin(\phi) \hat{y}
- \cos(\theta) \hat{z},
\end{align}
and
\begin{align}
\hat{p}_i=\sin(\theta_i)\cos(\phi_i) \hat{x}
+ \sin(\theta_i)\sin(\phi_i) \hat{y}
+ \cos(\theta_i) \hat{z}.
\end{align}
In these equations, $\phi_i$ and $\theta_i$ are the spherical coordinates of the $i$-th pulsar.
Finally, the noise spectral density is
\begin{equation}
\label{eq:nsd}
S_i=2\Delta t \sigma_i^2+S_{h,\text{rest}},
\end{equation}
where $2\Delta t \sigma_i^2$ is the contribution from the pulsar's white noise, and 
\begin{align}
S_{h,\text{rest}}=\frac{h_{c,\text{rest}}^2}{f} \frac{1}{12 \pi^2 f^2}
\end{align}
is an additional red noise term produced by all the other SBHBs present at the same frequency bin.
Consequently, $h_{c,\text{rest}}^2$ is obtained using Equation (\ref{eq:hc}), but the summation now includes the strain of all signals except for the \text{strongest} one, i.e. the one selected using Equation (\ref{eq:hcstrongest}).

If we neglect the evolution of the GW frequency during the observation time, the integral in Equation (\ref{eq:snralpha}) can be easily solved analytically.
In fact, the assumption that binaries do not evolve considerably (i.e. they are monochromatic signals) is a very accurate approximation, since the time-scales of the SBHB evolution is much longer than realistic observation times \citep[a discussion on this can be found in Section II.A.1 of][]{SesanaVecchio2010}.
Assuming monochromatic signals and after some algebra, Equation (\ref{eq:snralpha}) becomes
\begin{align}
\label{eq:sssnr_formula}
&\text{S/N}_i^2=\frac{A^2}{S_i 8\pi^3f^3}\left[ a^2 [F_i^+]^2 \left[ \Phi_T[1+2\sin^2(\Phi_0)] \right.\right. \nonumber\\
&\quad \left.+\cos(\Phi_T)[-\sin(\Phi_T)+4\sin(\Phi_0)]-4\sin(\Phi_0) \right] \nonumber\\
&+b^2[F_i^\times]^2\left[ \Phi_T[1+2\cos^2(\Phi_0)] + \sin(\Phi_T)[\cos(\Phi_T)-4\cos(\Phi_0)]\right] \nonumber\\
&\left.-2abF_i^+F_i^\times \left[ 2\Phi_T\sin(\Phi_0)\cos(\Phi_0)+\sin(\Phi_T)[\sin(\Phi_T)-2\sin(\Phi_0)]\right.\right.\nonumber\\
&\quad\left.\left.+2\cos(\Phi_T)\cos(\Phi_0)-2\cos(\Phi_0)\right]\right],
\end{align}
where
\begin{equation}
\Phi_T=2\pi fT.
\end{equation}
With the previous formulas we have all the necessary mathematical machinery to calculate the DP in Equation (\ref{eq:gammastotal}).

There is only one missing ingredient that we have not yet defined, which is the number of templates, $N$.
A careful computation of the independent number of templates in the search is a cumbersome task, and it is beyond the scope of this investigation.
A similar calculation has been performed by A. Petiteau in the context of the EPTA single source data analysis (Petiteau, private communication), by means of a stochastic template bank.
Petiteau covered the $(f,\theta,\phi)$ parameter space, defining independent templates with minimal overlap of 0.5.
He computed 4276 templates considering a volume in the frequency band 2 - 400 nHz and the whole sky.
Based on this calculation, we consider here $N=10^4$ templates as a reference number.
This choice is somewhat arbitrary, but we checked that our results have little dependence on $N$ across two order of magnitudes, in the range $10^3<N<10^5$.
 
\subsection{Pulsar arrays}
\label{sc:arrays}
We will analyse the ensemble of SBHBs in the simulated universes assuming different pulsar arrays:
\begin{enumerate}
\item The current IPTA. It consists of 49 pulsars with fixed sky positions.
The rms noise is of 0.5 $\mu s$ for 15 of the pulsars, 4 $\mu s$ for 25, and 5$\mu s$ for 9.
\item A simulated SKA1 array.
The exact design\footnote{\url{http://www.skatelescope.org}} of SKA1 is still under debate; furthermore, even when the detailed configuration of antennas and bandwidths is decided, it will not be clear how many low-noise millisecond pulsars will be detected over time.
For the purposes of this work, it is enough to assume a fixed amount of pulsars with a certain average rms noise.
In particular we assume that the SKA1 PTA will consist of 50 pulsars with noise around 100\,ns.
\item A simulated SKA2 array.
The characteristics of the SKA2 configuration are even more uncertain than those of SKA1.
Again, we adopt a simple approach and assume that the SKA2 PTA will be able to monitor 200 ms pulsars with noise around 50\,ns.
\end{enumerate}
In the SKA cases, the sky positions of the pulsars are chosen differently (and randomly) for each realisation of the Universe.
We do this, instead of just using one particular random distribution of pulsars over the sky, to avoid a possible bias in the results, that could be provoked by a particularly convenient or inconvenient distribution of pulsars.

As mentioned before, we consider GW signals in the frequency range $[f_\text{min}=T^{-1},f_\text{max}=\Delta t^{-1}]$ in frequency bins of width $\Delta f=T^{-1}$, and assume that the pulsar noise is described by a stationary Gaussian white process defined by a certain rms noise.
This oversimplistic approach does not take into account two complications:
\begin{enumerate}
\item The presence of residual uncorrelated red noise.
It is likely that all pulsars will show intrinsic red noise behaviour to some level (which has in fact already been detected in some of them); however, this is not always necessarily the case.
We therefore assume here no additional red noise, and defer the investigation of its effect to a future paper\footnote{This amounts to simply add an appropriate $S_{j,{\rm red}}$ to the noise power spectrum of each pulsar, yielding $P_j=2\sigma_i^2\Delta t+S_{j,{\rm red}}$.}.
\item The fitting for a pulsar timing model.
This is a step inherent to the timing process and cannot be hampered without affecting the final results.
Most importantly, the timing model fits out a quadratic function $g(t)=a+bt+ct^2$ encoding the pulsar spin-down.
This will partially absorb the red spectrum imprinted by the GWB in each individual pulsar, affecting the detection capability of the array.
As shown by \cite{MooreEtAl2015}, this will likely affect the sensitivity in the lowest couple of frequency bins (see Figure 1 in that paper).
\end{enumerate}

Our main goal is not to make accurate predictions about {\it when} a PTA will detect the first GW signal, but rather about {\it what} kind of signal it will be.
The answer to this second question is also somewhat dependent on the treatment of the timing model, but to a much lesser extent.
We plan to implement consistently the effect of both timing model and additional red noise in the future.
In this first paper we will show results assuming the whole frequency domain, $[T^{-1},\Delta t^{-1}]$, removing the lowest frequency bin, $[2T^{-1},\Delta t^{-1}]$, and removing the two lowest frequency bins, $[3T^{-1},\Delta t^{-1}]$.
A careful implementation of the effect of the timing model is expected to lie between the two latter cases, i.e., removing the lowest and the two lowest frequency bins.

\section{Results}
\label{sc:results}
Having described all the relevant methods, we now turn to the discussion of our main results.
For each of the 234000 ensembles of SBHBs, we compute $h_c$, using Equation (\ref{eq:hc}), progressively increasing the observation time $T$ from 1 to 30\,yr in steps of 1\,yr for the IPTA, and from 0.5 to 10\,yr in steps of 0.5\,yr for the SKA1 and SKA2 arrays.
For each value of $T$ we then compute the DP of the GWB according to the B-statistic using Equation (\ref{eq:DP}) and the S/N from Equation (\ref{eq:snrb_bin}).
We also identify the loudest sources at each frequency bin and compute the DP of a single source and its S/N according to Equations (\ref{eq:gammastotal}) and (\ref{eq:snrs}), respectively.

\subsection{Detection probability as a function of time}
\begin{figure}
\begin{tabular}{c}
\includegraphics[scale=0.9,clip=true,angle=0]{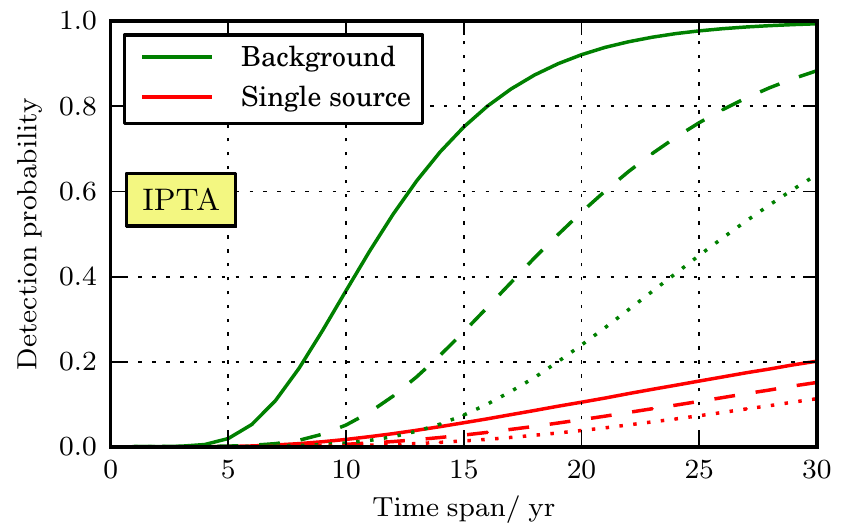}\\
\includegraphics[scale=0.9,clip=true,angle=0]{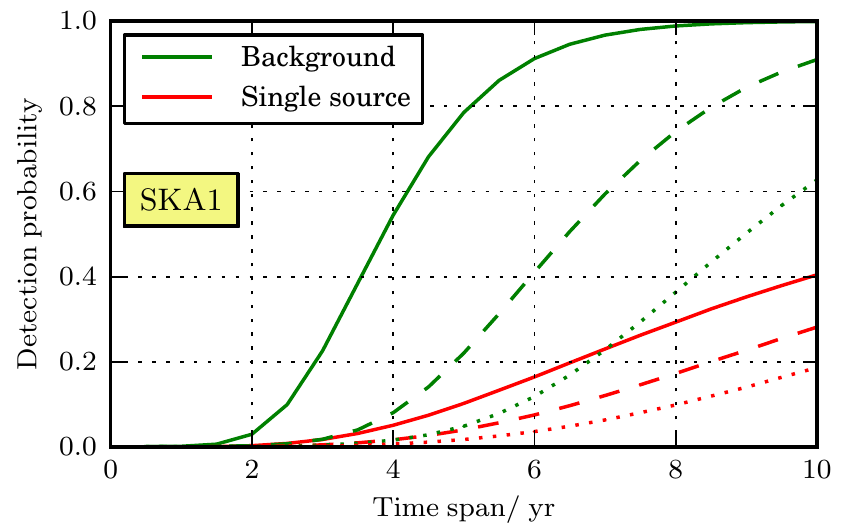}\\
\includegraphics[scale=0.9,clip=true,angle=0]{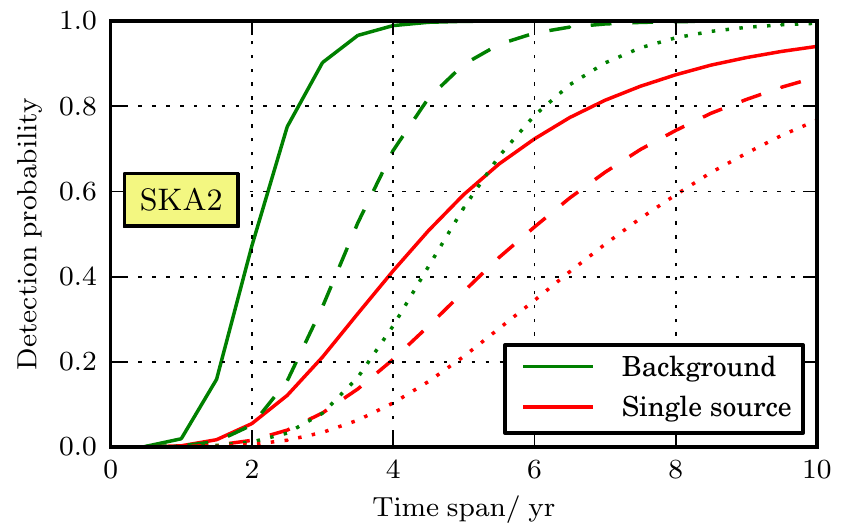}\\
\end{tabular}
\caption{Detection probability (averaged over all realisations of the ensemble of SBHBs) versus observing time of a GWB (green lines) and a single source (red lines) assuming the IPTA (upper panel) SKA1 (middle panel) and SKA2 (lower panel).
Solid lines show DPs integrated over all the frequency range $[T^{-1},\Delta t^{-1}]$, whereas the dashed (dotted) lines show the result when the lowest frequency bin (two lowest bins) are removed from the calculation.}
\label{fig:dp_all}
\end{figure}

Although addressing {\it when} we will detect GWs with PTAs is not the main goal of this work, the first natural outcome of our calculation is the DP of each class of sources as a function of time: $\gamma_\text{B}$ for a GWB, and $\gamma_\text{S}$ for an individual source.
This is presented in Figure \ref{fig:dp_all} for the three investigated arrays.
In each panel, we show the result for a specific PTA, considering the whole signal and progressively subtracting the lowest frequency bins.

It is clear from Figure \ref{fig:dp_all} that the subtraction of the lowest frequency bins has a larger impact on the DP of the GWB than on the DP of single sources.
This is because there are more binaries emitting at lower frequencies, and thus the GWB becomes stronger, and more likely to be detected, at the lowest frequency bins.
For single sources, the trend is different.
On the one hand, at very low frequencies there are more binaries, but they are hardly resolvable, i.e. the GWs produced by all other binaries act as a noise that masks the signal of an individual source.
On the other hand, at high frequencies, although single sources are easier to resolve (since there are fewer binaries emitting at similar frequencies), it is less likely to find one emitting strong enough GWs to be detectable.
There is thus a particular range of frequencies (not necessarily the lowest or the highest frequency bins) where single sources are more easily detectable.

The green curves in the upper plot of Figure \ref{fig:dp_all} show the predicted DP of a GWB for the IPTA.
The current array has been running for $\approx10\,$yr, which corresponds to a DP of $\approx37\%$ assuming pure white noise (solid green curve).
However, the DP at 10\,yr goes down to $\approx5\%$ and $<1\%$ when the lowest and the two lowest frequency bins are taken out of the computation.
Another way to put this is by looking at the $50\%$ DP: this probability is reached in approximately 2\,yr, 9\,yr, and 16\,yr from now, depending on the cut-off on the low frequency bins.

The DP for individual sources with the IPTA (red curves in the upper plot of Figure \ref{fig:dp_all}) is, on the other hand, always relatively small, reaching values around $10-20\%$ after 20 years from now.
This should not be taken as an indication against the development of single source searches.
Firstly, we adopted a fairly conservative criterion for the detection of single sources, placing a threshold of FAP $\alpha=0.1\%$; it could be possible that the signal looked like a collection of relatively faint hotspots in the sky, not classified as single sources in our computation, but also not similar to a GWB.
A search for multiple single sources \citep{BabakSesana2012} might be more efficient than a search for a GWB in such situation.
Secondly, these predictions assume the current array {\it without} any future improvement.
Adding more pulsars and, most importantly, reducing the rms of individual pulsars will result in a much higher chance of detecting a single source, which is what we expect as we enter the SKA era.
Indeed, this is shown in the middle and lower panels of Figure \ref{fig:dp_all} (note that the x-axis now runs to 10 year only).
With the full SKA PTA (described in Section \ref{sc:arrays}) the detection of a GWB will be almost certain within 5 years, whereas single sources will be detected with a probability higher than $80\%$ after 10 years.
The higher probability of detecting individual sources stems from the lower rms expected from individual pulsars in the SKA era.
The PTA will start to `hit the signal' at much higher frequencies ($f>10^{-8}\,$Hz), where it is also likely to see individual sources and not only an unresolved GWB. We notice, nonetheless, that these numbers  are quite less optimistic that those quoted in \cite{JanssenEtAl2015}. This is because, in that paper, both the GWB and single sources are said to be detectable if the produced S/N is larger than 4. However, it turns out that this is not sufficiently stringent for claiming a confident detection of a single source. One should consider the results of this work as an update to those claimed in \cite{JanssenEtAl2015}, keeping in mind some important caveats related to the possible resolution of multiple sources that we will discuss in Section \ref{sc:discussion}.

Let us stress once again that when calculating the DP for the SKA, we neglect previous IPTA data.
One could achieve a more accurate prediction by adding millisecond pulsars to the IPTA gradually, until reaching the SKA1 configuration.
This may be considered for a future work, whereas in this paper we attempt to focus on the detection capabilities of the current IPTA, and compare it with a PTA constructed with the SKA.

\subsection{Signal-to-noise ratio}
\label{subsc:snr}
\begin{figure}
\includegraphics{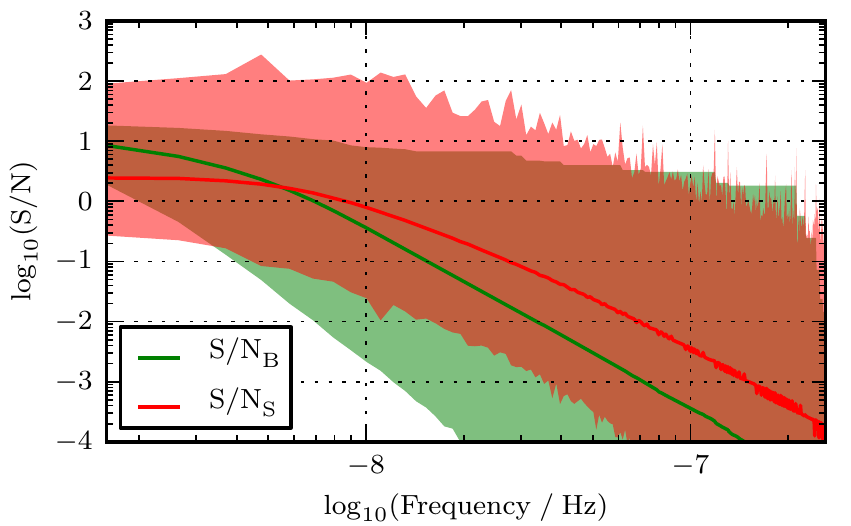}
\caption{$\text{S/N}_\text{S}$ and $\text{S/N}_\text{B}^{f_i}$ versus observed GW frequency. The red area contains the $\text{S/N}_\text{S}$ values of the strongest binaries in each frequency bin, for all of our realisations of the SBHB population; the red line gives the average $\text{S/N}_\text{S}$ over all realisations. The green area contains the cumulative contributions to $\text{S/N}_\text{B}$ above each particular frequency bin, $\text{S/N}_\text{B}^{f_i}$, defined in Equation (\ref{eq:snrbfi}); the green line gives the mean $\text{S/N}_\text{B}^{f_i}$ at each bin. Thus, the left-most point of the green line gives the ensemble average $\text{S/N}_\text{B}$.}
\label{fg:snr}
\end{figure}

\begin{figure}
\includegraphics[scale=0.9,clip=true,angle=0]{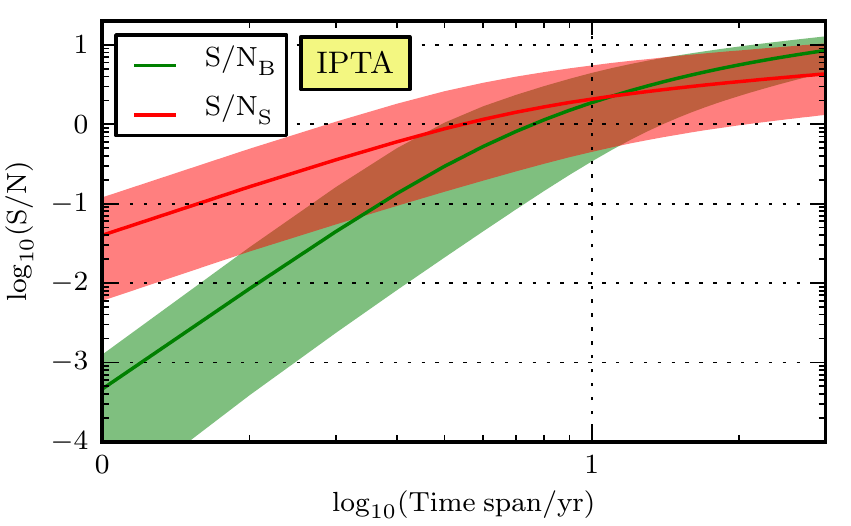}
\includegraphics[scale=0.9,clip=true,angle=0]{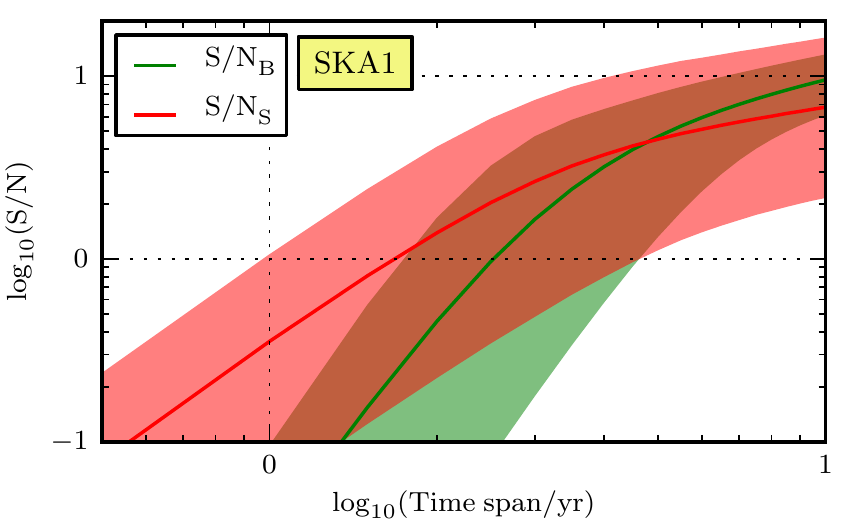}
\includegraphics[scale=0.9,clip=true,angle=0]{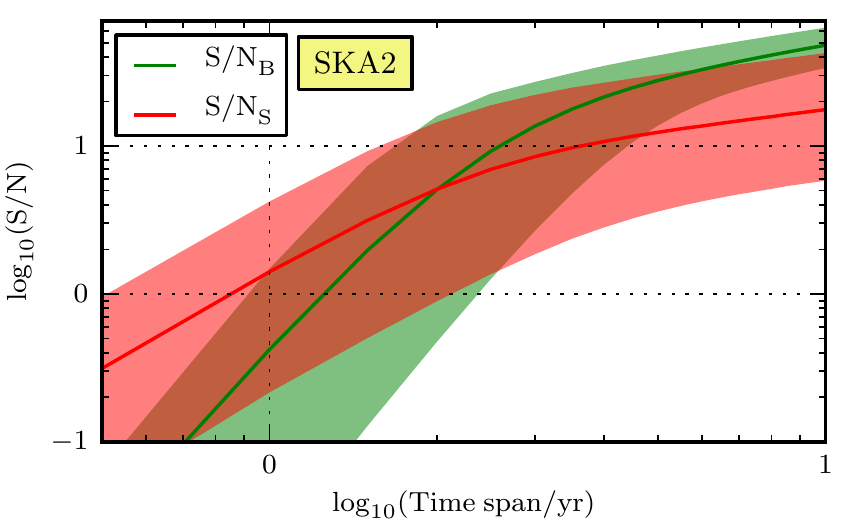}
\caption{S/N versus observation time, for an IPTA (top), SKA1 (central), and SKA2 (bottom).
The red area covers the values between the 5th and 95th percentiles of $\text{S/N}_\text{S}$ of all realisations; the red line is the average over realisations.
The green area covers values between the 5th and 95th percentiles of $\text{S/N}_\text{B}$, and the green line is the average.}
\label{fg:snrevol}
\end{figure}

As explained in Section \ref{sc:models}, the probability of detecting a GW signal is evaluated in this paper using the DP instead of the S/N, which is more commonly used in the related literature.
We now explore the properties of the S/N of the two types of signals.

This is shown in Figure \ref{fg:snr} for the IPTA; there we compare $\text{S/N}_\text{B}$ with $\text{S/N}_\text{S}$, calculated as explained in Sections \ref{sc:gwb} and \ref{sc:single} respectively, assuming an integration time of 30yr, and a frequency bin of $\Delta f=[30\,\text{yr}]^{-1}$.
The red area contains the $\text{S/N}_\text{S}$ of the strongest binary at a given frequency bin, in each of the realisations; the red line is the mean $\text{S/N}_\text{S}$ over all realisations.
The green area, instead, contains the cumulative contribution to $\text{S/N}_\text{B}$ (whose formula will be derived in Appendix \ref{app:stats}) from frequencies above a particular frequency bin, i.e.
\begin{align}
\label{eq:snrbfi}
&\text{S/N}_\text{B}^{f_i}=\nonumber\\
&\left[ 2 \sum_{f_i}^{f_\text{max}} \sum_{i=1}^M \sum_{j>i}^M \frac{\Gamma_{ij}^2 S_h^2}{P_iP_j+S_h[P_i+P_j]+S_h^2[1+\Gamma_{ij}^2]}\right]^{1/2}.
\end{align}
The green line in the figure is, again, the average over all realisations.
Note that the values of $\text{S/N}_\text{B}^{f_i}$ at each frequency bin are irrelevant; it is the overall values of $\text{S/N}_\text{B}$ (which are the left-most points in the green area) that are actually meaningful.
On the other hand, at each frequency bin we have independent single sources, some of which might be bright enough to be resolvable.
Therefore all values of $\text{S/N}_\text{S}$ are equally relevant.
We point out that, at $f\gtrsim 10^{-8}\,$Hz, the typical S/N of the brightest singles sources is indeed higher than the cumulative S/N of the GWB down to the same frequency, whereas it flattens out at $f\lesssim10^{-8}$Hz, where there are more and more sources per frequency bin, and it is difficult for a single SBHB to dominate the signal.

The evolution of the S/N as a function of the observation time is shown in Figure \ref{fg:snrevol}, for IPTA, SKA1 and SKA2.
The S/N presents similar features for both classes of sources, behaving as a double power law; however, their detailed evolution is different, and the two curves end-up intersecting at some point.
The S/N produced by a GWB follows the trend already predicted by \cite{SiemensEtAl2013}, which is not surprising, given that the formula they use to calculate the S/N is almost identical to ours (numerically, the difference is negligible), in Equation (\ref{eq:snrb_bin}).
One thing that is worth noticing is that the specific value at which the S/N flattens out for long observing times does only depend on the number of pulsars in the array.
In the IPTA case, where 15 pulsars give a major contribution to the detection statistic (having a rms noise which is 8 times better than the others) the turnover starts already at S/N$\approx1$, whereas in the SKA2 case, where 200 pulsars equally contribute to the array, the turnover starts only at S/N$\approx10$.
In practice we have a turnover by S/N $\propto M^{1/2}$, where $M$ is the number of pulsars.
Therefore, having an array formed by many pulsars with decent precision might be a better strategy to detect a GWB than having few ultra-precise pulsars \citep[a point that has also been made by][]{SiemensEtAl2013}.

The S/N of single sources grows initially faster than quadratically in time, because by accessing lower and lower frequencies (but still higher than $10^{-8}\,$Hz, where SBHBs are still quite sparse) there is a larger chance to detect a new bright individual source.
However, after enough observation time, when frequencies lower than $10^{-8}\,$Hz start to be accessible, individual sources with high S/N become less likely to be found. At that stage, the S/N of an individual source grows as $T^{1/2}$, as suggested by Equation (\ref{eq:snralpha}).

By inspecting the upper plot in Figure \ref{fg:snrevol}, one could claim that the detection of a single binary is currently (by $\approx10\,$yr of observing time span for the IPTA) more likely than that of a background, since the average $\text{S/N}_\text{S}$ is larger than $\text{S/N}_\text{B}$.
Moreover, if one assumes that a detection is achieved as soon as the S/N surpasses a certain threshold \citep[for example 4, as in][]{JanssenEtAl2015}, one would also claim that detecting a single SBHB is more likely than detecting a GWB for the two SKA configurations considered.
However, both claims would be wrong; as we mentioned in Section \ref{sc:models}, the detectability should be evaluated in terms of the DP, and, as shown in Figure \ref{fig:dp_all}, the probability of detecting a GWB is in all cases (for all PTAs and observing times considered) larger than that of detecting a single SBHB.

\subsection{What will be detected first?}
\label{sc:binaries}

\begin{figure}
\begin{tabular}{c}
\includegraphics[scale=0.9,clip=true,angle=0]{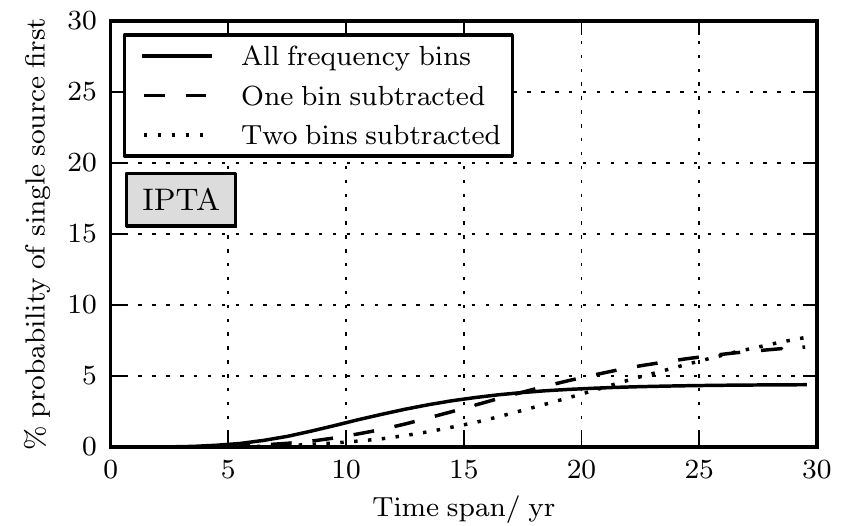}\\
\includegraphics[scale=0.9,clip=true,angle=0]{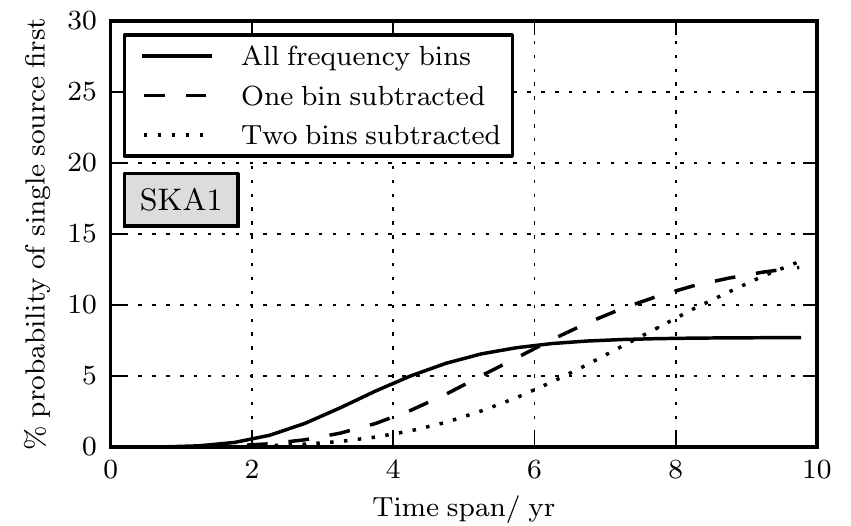}\\
\includegraphics[scale=0.9,clip=true,angle=0]{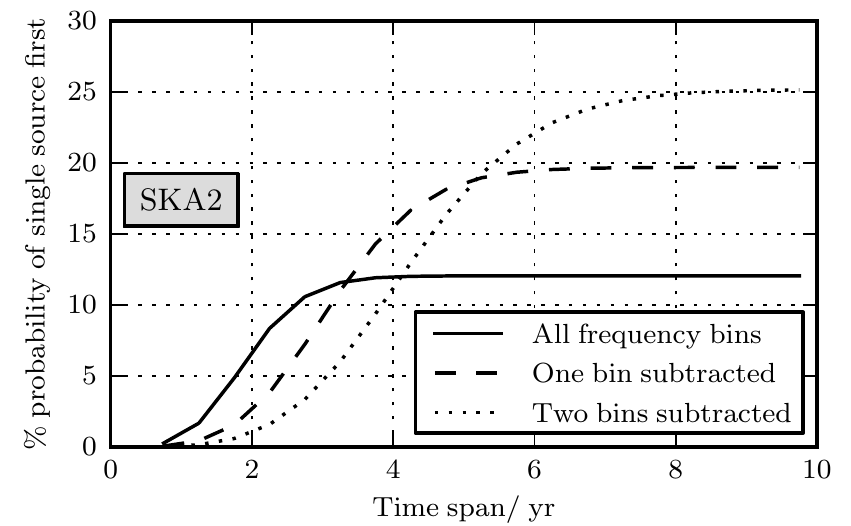}\\
\end{tabular}
\caption{Cumulative probability of detecting a single resolvable source {\it before} a GWB versus observing time, for the IPTA (top panel), SKA1 (central panel) and SKA2 (bottom panel).
In each panel, solid lines show DPs integrated over all the frequency range $[T^{-1},\Delta t^{-1}]$, whereas the dashed (dotted) lines show the result when the lowest bin (two lowest bins) are removed from the calculation.}
\label{fig:dp_first}
\end{figure}

Having inspected the DP of each class of sources, we can now tackle our original problem: which kind of signal is more likely to be detected first by a PTA?
To answer this question we first need to know the probability of a GWB being detected between times $t$ and $t+dt$; let us call this probability $p_B(t)dt$.
The probability of detecting a GWB after a time $t$ would then be
\begin{equation}
\gamma_B(t)=\int_0^t p_B(t')dt',
\end{equation}
where $\gamma_B$ is the DP of a GWB, given by Equation (\ref{eq:DP}).
Similarly, the probability $p_S(t)dt$ of a single source being detected between times $t$ and $t+dt$, would be related to the DP of a single source (in Equation (\ref{eq:gammastotal})) by
\begin{equation}
\gamma_S(t)=\int_0^t p_S(t')dt'.
\end{equation}
We can numerically obtain the probability density functions $p_B(t)$ and $p_S(t)$ from $\gamma_B(t)$ and $\gamma_S(t)$, respectively.
Once those functions are known, we can calculate the probability that a single source is detected between $t$ and $t+dt$ given that a GWB has not been detected at any time before $t+dt$; this would be given by $[1-\gamma_B(t)]p_S(t)dt$.
Finally, we can define the function
\begin{equation}
P_S(T)=\int_0^T [1-\gamma_B(t)]p_S(t)dt,
\end{equation}
which gives the cumulative conditional probability over time of a single source being detected after a time $T$, given that a GWB has not been previously detected.
This is the quantity that we need to evaluate in order to answer the central question of this paper.

We compute $P_S(T)$ as a function of the observation time $T$ for the different arrays considered in this work; the result is shown in Figure \ref{fig:dp_first}.
There we see that, if the IPTA array keeps going for other 20 years, there is only about a 4-8\% probability that the first claimed detection is of a single source.
Thus, the first detected PTA signal will most likely be a GWB or an incoherent superposition of multiple sources.
Things will change significantly (but not dramatically) in the SKA era.
An SKA1-type PTA will have a 8-13\% chance to detect a single source first, a percentage that might increase to about 12-25\% in the SKA2 phase.
Note that the aforementioned probabilities depend on the treatment of the lowest frequency bin.
In particular, when all low frequency bins are considered, the DP of the GWB becomes much larger than that of single sources, and hence the probability of detecting a single source before a GWB becomes lower and saturates earlier (solid curves in Figure \ref{fig:dp_first}).
To be conservative, we assume that the lowest frequency bin has a smaller contribution in the signal build-up, because of the timing model fitting; then, omitting the results in which all frequency bins are considered, we can confidently say that the chance to detect a single source before a GWB lies between 7\% and 25\%, depending on the array configuration.

\subsection{Properties of the first detectable single source}
\label{sc:binaries}

\begin{figure*}
\begin{tabular}{ccc}  
\includegraphics[scale=0.63,clip=true,angle=0]{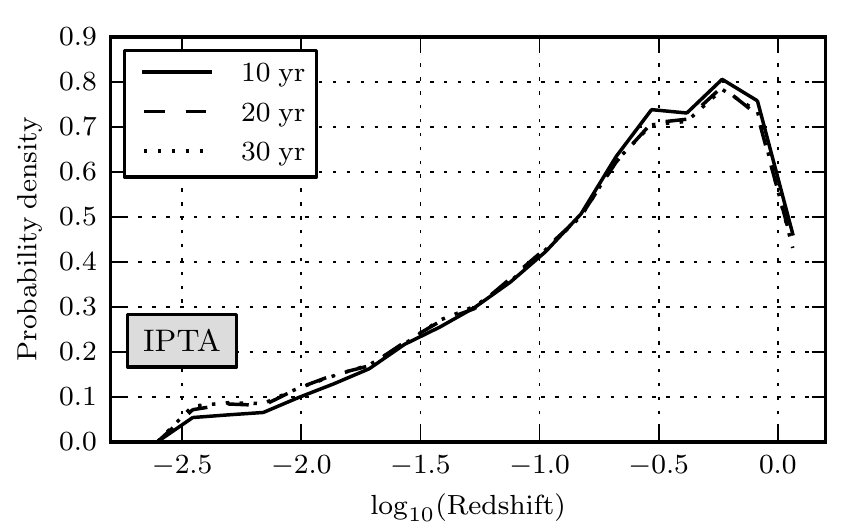}&
\includegraphics[scale=0.63,clip=true,angle=0]{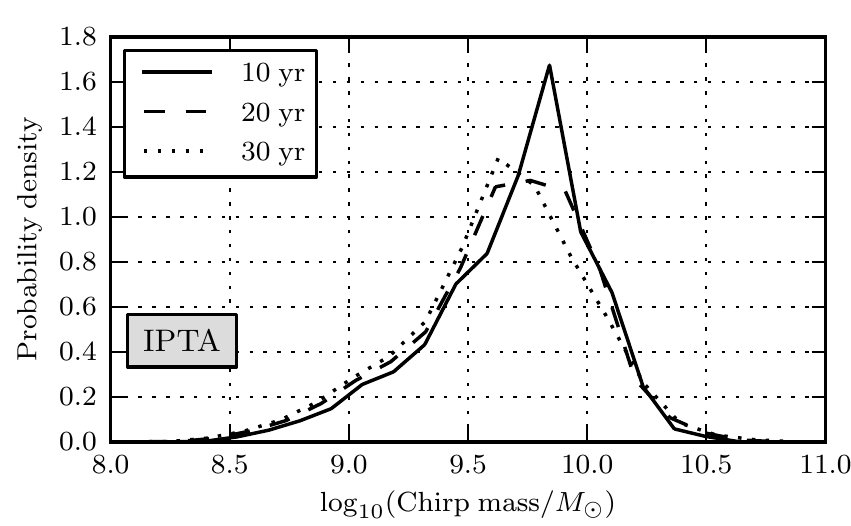}&
\includegraphics[scale=0.63,clip=true,angle=0]{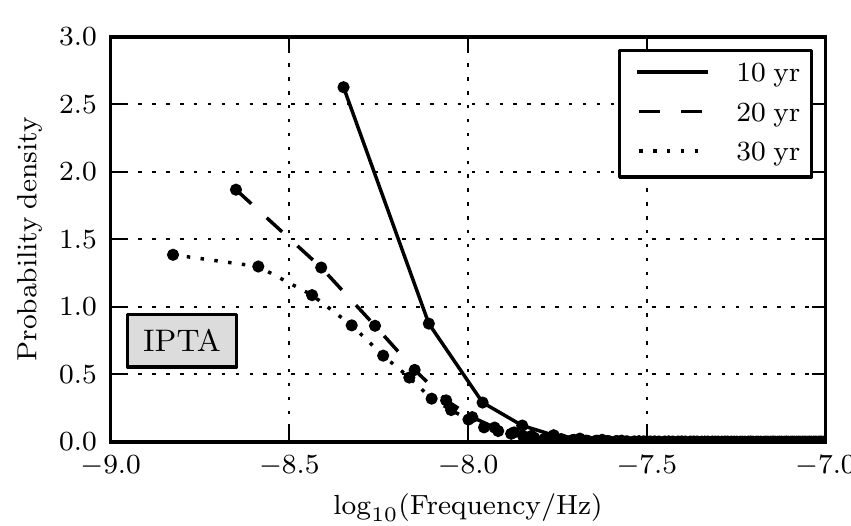}\\
\includegraphics[scale=0.63,clip=true,angle=0]{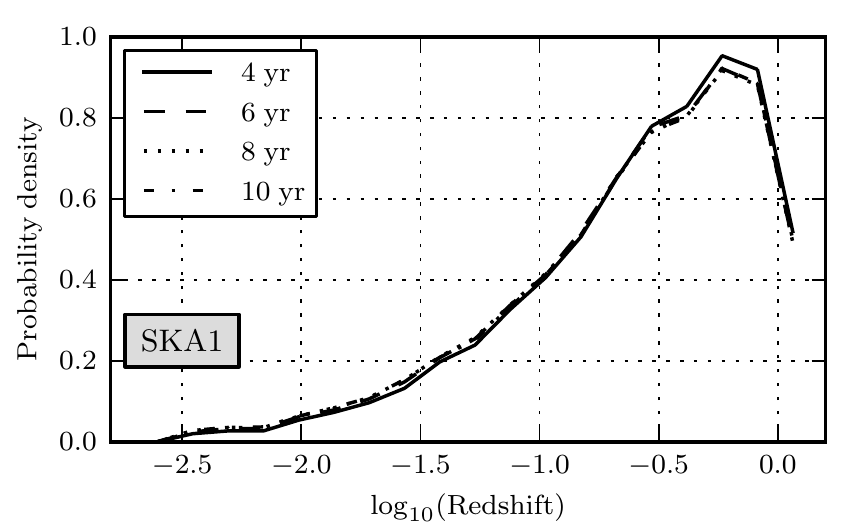}&
\includegraphics[scale=0.63,clip=true,angle=0]{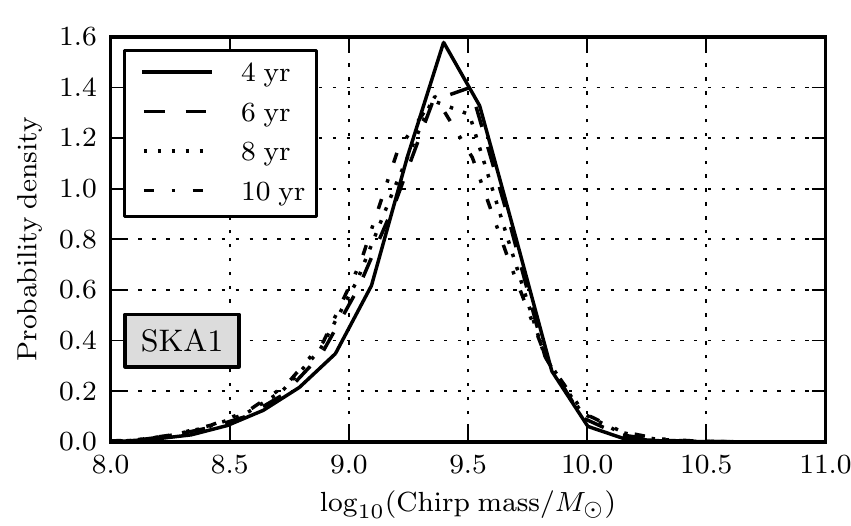}&
\includegraphics[scale=0.63,clip=true,angle=0]{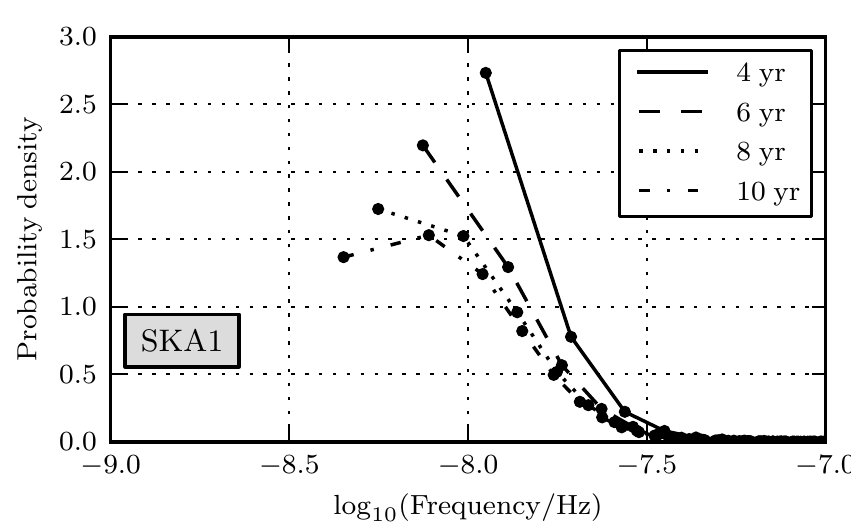}\\
\includegraphics[scale=0.63,clip=true,angle=0]{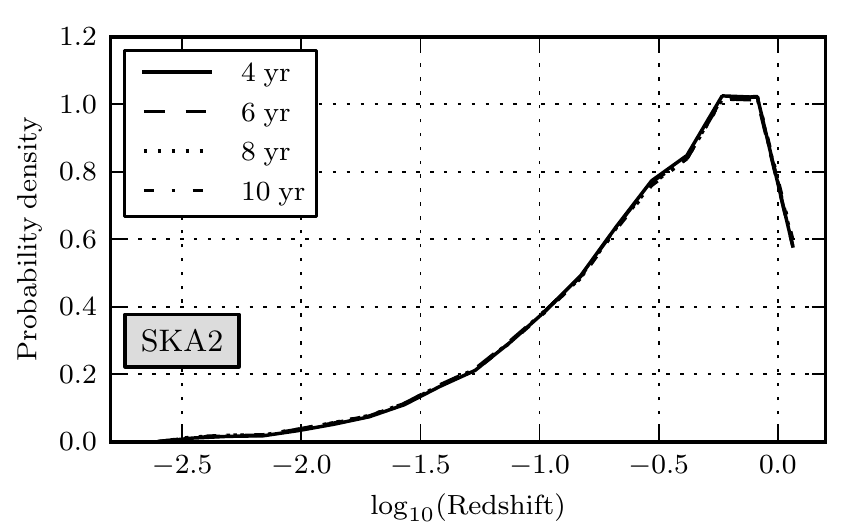}&
\includegraphics[scale=0.63,clip=true,angle=0]{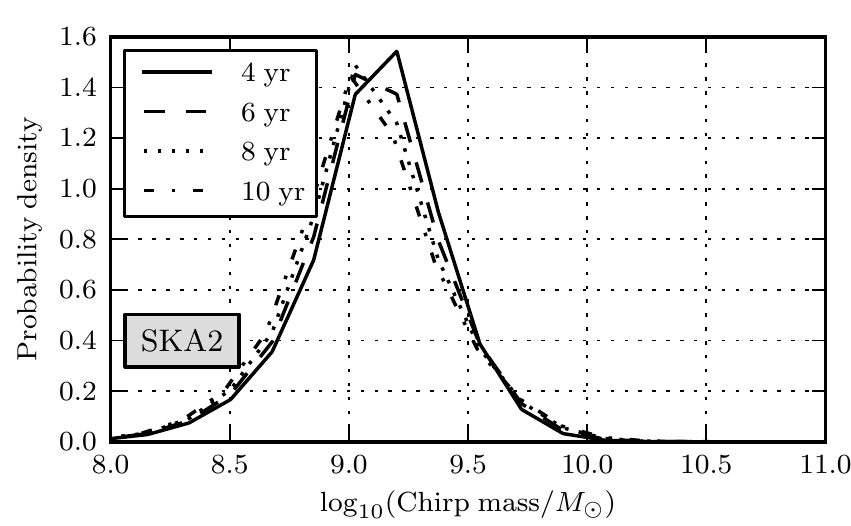}&
\includegraphics[scale=0.63,clip=true,angle=0]{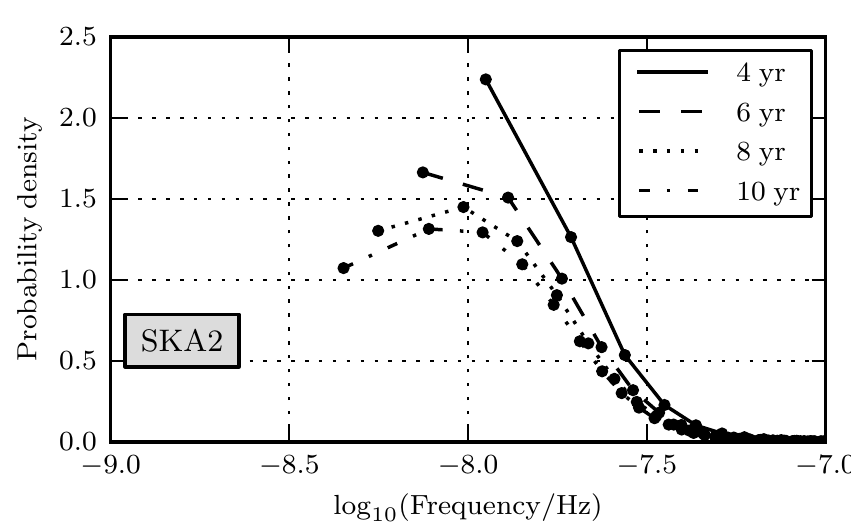}\\
\end{tabular}
\caption{Properties of the individual SBHBs that are most likely to be detected with the current IPTA (upper row plots), SKA1 (central row) and SKA2 (lower row).
The different columns of plots show, from left to right, the probability density function of redshift, chirp mass, and GW observed frequency of the individual SBHBs.
The black points in the graphs on the right column are located at the arithmetic mean of the frequency bins.}
\label{fig:ss_all}
\end{figure*}

Even though single sources will most likely be detected after a GWB, they will eventually pop-up in PTA data.
It is therefore interesting to study their properties in terms of mass, redshift and GW frequency.
We summarise this information in Figure \ref{fig:ss_all} for all the investigated arrays and different observation times.
All frequency bins were taken into account when constructing this figure (even the lowest ones).
To construct these distributions, we took all the realisations of the SBHB population and considered, at any observation time, the brightest source in each frequency bin, weighting their contribution by their DP.

The properties of the first detectable single sources do not depend strongly on the assumed PTA.
The vast majority of the sources are massive, nearby binaries, clustering in the redshift range $0.3-0.7$, regardless of the PTA properties.
In terms of chirp mass, the more sensitive the array, the less massive the first detectable binaries would be.
Therefore, although the IPTA has a chance to detect binaries with ${\cal M}>10^{9.5}\msun$, SKA2 will observe many more systems with ${\cal M}\approx10^{9}\msun$.
The rms noise of the pulsars in the array also has an impact in the frequency distribution of the systems (right panels).
In a putative IPTA scenario, the higher DP is always at the lowest frequency, despite the fact that more sources contribute to the signal in that range, and the inherent probability of having a single source sticking out is smaller.
This is simply because the individual pulsar rms in the IPTA are not good enough to allow efficient detection at high frequencies, and sources must be extremely bright to be observed there.
Conversely, in both SKA scenarios (central and lower rows of plots), the single source DP peaks around $10^{-8}$Hz, even if the observation time is extended to 10 years and lower frequencies are accessible.
This is simply because the array sensitivity now allows the detection of individual sources also at higher frequencies, making clear that the presence of these systems becomes intrinsically less probable at progressively lower frequencies. 

\subsection{Pulsar term of the first single source}
\label{sc:pulsar_term}

\begin{figure*}
\begin{tabular}{ccc}  
\includegraphics[scale=0.63,clip=true,angle=0]{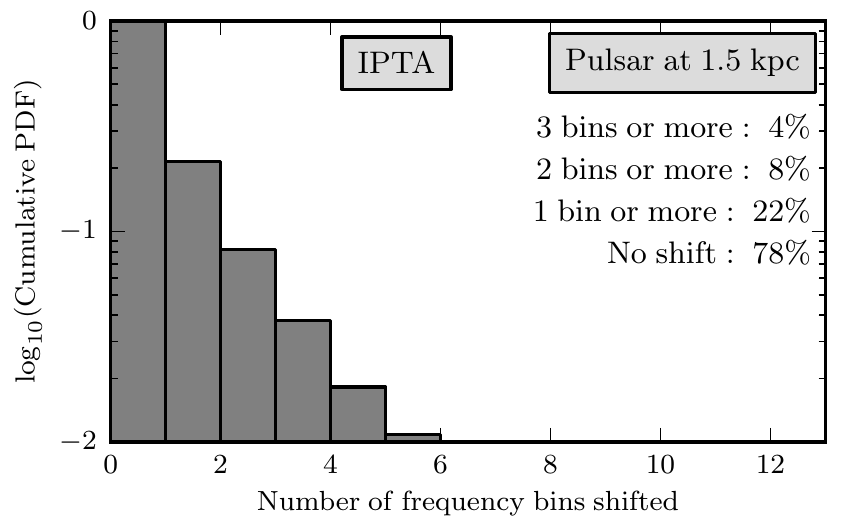}&
\includegraphics[scale=0.63,clip=true,angle=0]{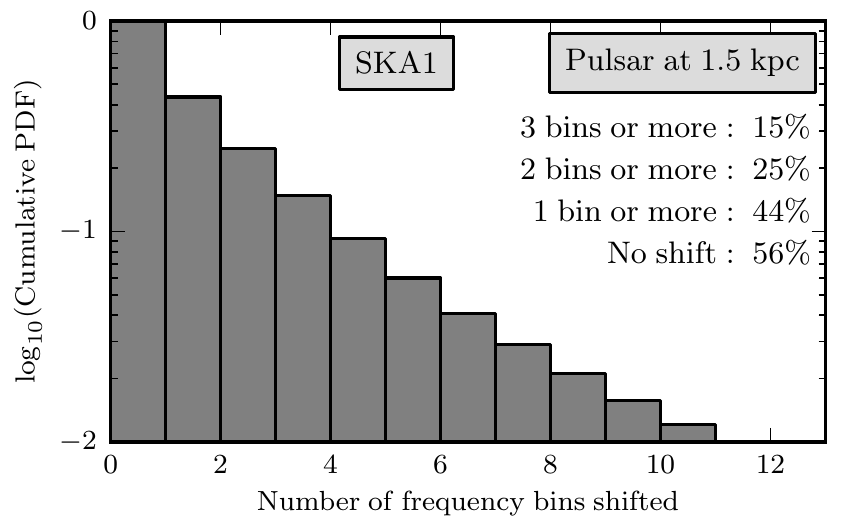}&
\includegraphics[scale=0.63,clip=true,angle=0]{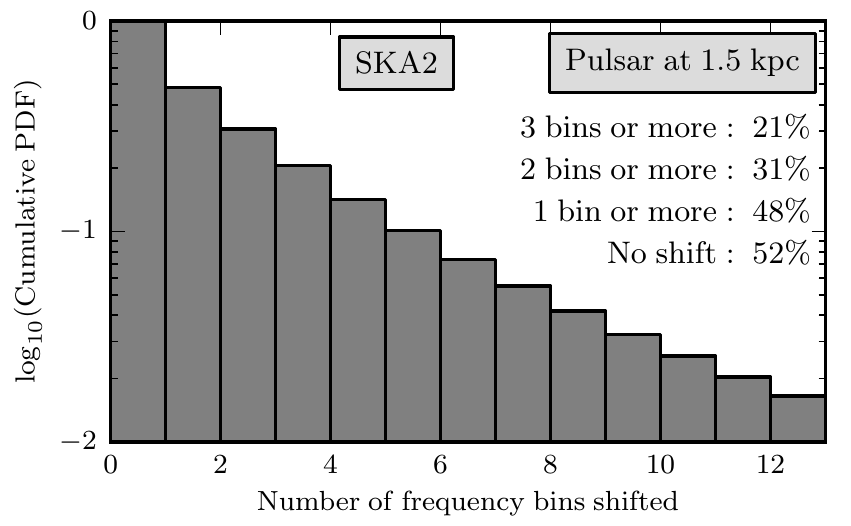}\\
\includegraphics[scale=0.63,clip=true,angle=0]{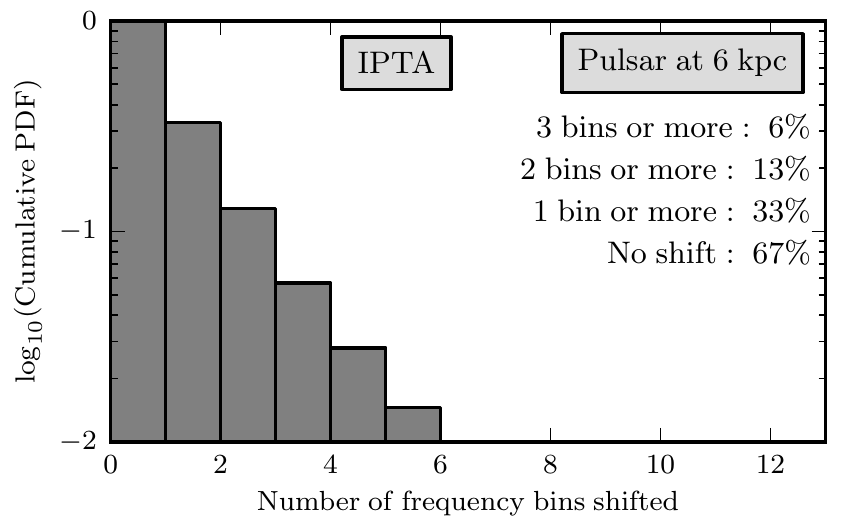}&
\includegraphics[scale=0.63,clip=true,angle=0]{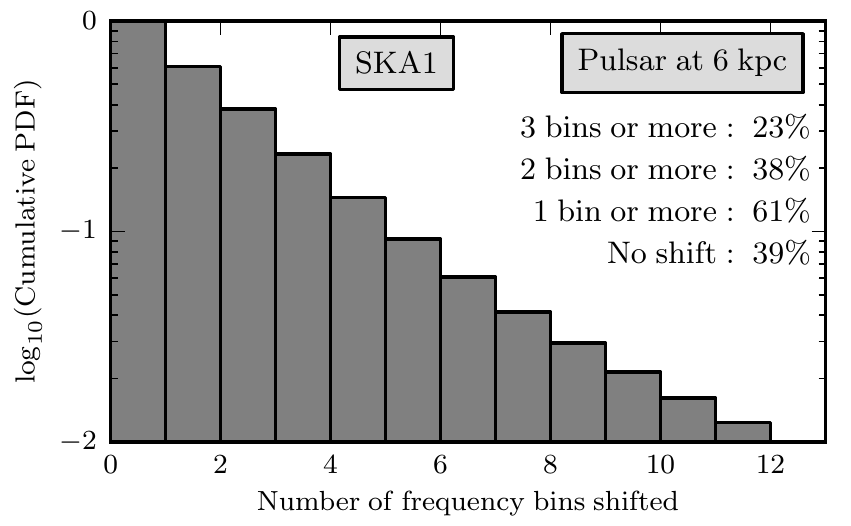}&
\includegraphics[scale=0.63,clip=true,angle=0]{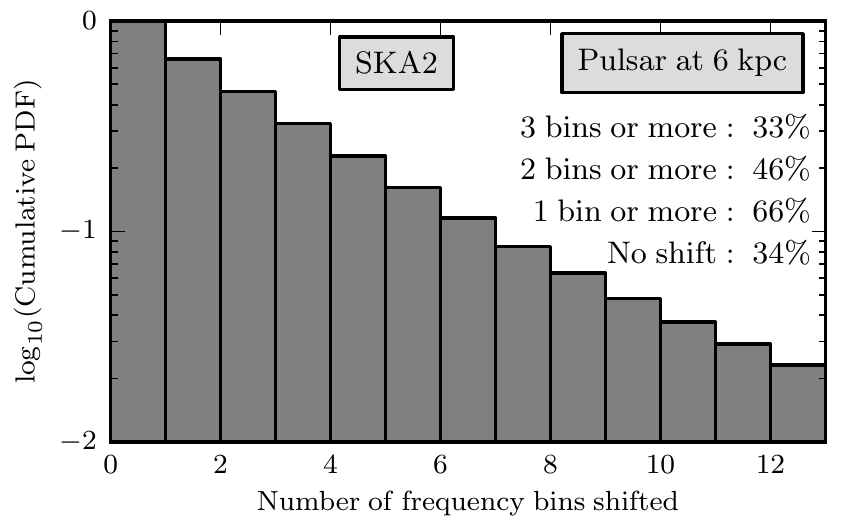}\\
\end{tabular}
\caption{Histogram of the frequency shift between the pulsar and Earth terms, in units of the frequency bin $T^{-1}$), where $T$ is the observing time (assumed 10\,yr in all cases), for the current IPTA (left panels), SKA1 (central panels) and SKA2 (right panels).
The upper plots assume a typical observed time delay consistent with the mean estimated pulsar distance in the current IPTA ($\approx1.5\,$kpc). The lower plot assumes a typical time delay consistent with the maximum pulsar distance in the current IPTA ($\approx6\,$kpc).}
\label{fig:SS_all_bins}
\end{figure*}


The GWs arriving at our galaxy produce a deformation of the space-time metric that can be observed in two ways: via the Earth term and via pulsar terms.
The Earth term affects the ToAs of the pulses of all pulsars at the moment when the GWs reach Earth.
On the other hand, a pulsar term is originated when the GWs reach any of the pulsars; the distorted pulses emitted by that pulsar are observed by our telescopes after they travel all the way from the pulsar to Earth.
Thus, there is a delay between the Earth term and any observable pulsar term, of hundreds or thousands of years, depending on the distance to the pulsar and its relative angle with respect to the GW source.
Consequently, it is possible to observe GWs of a single source at two different stages of the binary's lifetime, differing hundreds or thousands of years.
The question we address now is, does the GW frequency differ significantly between the Earth and pulsar terms?
This is an interesting topic that has consequences in the development of data analysis algorithms targeting continuous GW sources.

Let us call $f^\text{P}$ the observed GW frequency of a pulsar term, and $f^\text{E}$ the observed GW frequency of the Earth term.
We assume that $z$ is the redshift of the GW source, which can be safely considered invariant during time-scales of thousands of years.
Then, at the time when the SBHB emitted those waves, the pulsar and Earth terms had frequencies $f_e^\text{P}=[1+z]f^\text{P}$ and $f_e^\text{E}=[1+z]f^\text{E}$ \citep[using a similar nomenclature as in][]{Rosado2011}.
The interval of time that the SBHB needed to evolve between the emission of those waves is given by
\begin{equation}
\label{eq:deltate}
\Delta \mathcal{T}_e=\frac{5c^5}{256\pi^{8/3}[G\mathcal{M}]^{5/3}}\left[[f_e^\text{P}]^{-8/3}-[f_e^\text{E}]^{-8/3}\right].
\end{equation}
That interval of time would be today observed as $\Delta \mathcal{T}=\Delta \mathcal{T}_e[1+z]$.
Introducing this in Equation (\ref{eq:deltate}) and rearranging, we get
\begin{equation}
f^\text{P}=\left[\frac{\Delta \mathcal{T} 256 \pi^{8/3}[G\mathcal{M}[1+z]]^{5/3}}{5c^5}+[f^\text{E}]^{-8/3}\right]^{-3/8}.
\label{eq:deltaf}
\end{equation} 
Equation (\ref{eq:deltaf}) allows us to quantify the frequency of the pulsar term ($f^p$) once the chirp mass, redshift and Earth term frequency ($f^E$) of a putative resolvable source (all information we get directly from our Monte Carlo realisations of the Universe) are known. The only other quantity we need to fix is the delay $\Delta \mathcal{T}$ between the arrival of the Earth term and the pulsar term, which depends on the typical distances of the observed pulsars to the Earth, that we next quantify.

The most distant pulsar in the current IPTA is at $\sim 6\,$kpc.
The maximum delay between pulsar and Earth terms for this pulsar would be of $\sim12\,$kpc$/c$, which is twice the light travel time between that pulsar and Earth.
This delay would be achieved if the SBHB was located exactly at the opposite sky location of the pulsar.
We calculate $f^\text{P}$ of all binaries in each realisation of the Universe, assuming $\Delta \mathcal{T}=12\,$kpc$/c$, and also $\Delta \mathcal{T}=3\,$kpc$/c$, which would be a typical delay for average IPTA pulsars (the mean estimated distance to IPTA pulsars is $\approx1.5$).
Then we obtain the frequency shift between the pulsar and the Earth term, $[f^\text{E}-f^\text{P}]/\Delta f$, where $\Delta f=T^{-1}$, and $T$ is the observing time needed to detect the SBHB.

The distribution of frequency shifts of the first detectable binary for all the considered PTAs is shown in Figure \ref{fig:SS_all_bins}, assuming $\Delta \mathcal{T}=3\,$kpc$/c$ (upper plot) and $\Delta \mathcal{T}=12\,$kpc$/c$ (lower plot).
All histograms are weighted by the DP of the binaries, so that binaries with larger probability of being detected are more representative than those with smaller DP.
By looking at the figure one can conclude that a significant evolution between pulsar and Earth term is not likely; among binaries detected in the near future with the IPTA, a shift of at least 1 bin has a probability of 22\% for average pulsars (at around 1.5\,kpc), and to 33\% for a pulsar at 6\,kpc.
These percentages increase as we move into the SKA era; for an SKA2 arrays the odds of having a sizeable frequency shift between the pulsar and the Earth terms become higher than 50\%, due to the fact that single sources are detected at higher frequencies, where their chirping becomes progressively faster.
All in all, we can conclude that frequency-shifting and strictly monochromatic individual sources are roughly equally likely to be detected.

\section{Discussion and conclusions}
\label{sc:discussion}

The results shown in the previous section have a number of important implications for the future design of PTA observations and data analysis pipelines.

\begin{enumerate}
\item We showed that a GWB is more likely to be detected than any resolvable single source, which is consistent with other recent predictions \citep{RosadoSesana2014,RaviEtAl2015}.
However, this result should not discourage the development of single source detection pipelines for at least two reasons.
First, there is a 5-to-25\% probability of detecting a single source first, which is not negligible; in particular, if the IPTA sensitivity will prove insufficient, expected timing improvement in the SKA era will result in an increase of the odds of detecting a single source.
Second, throughout the paper we adopted a conservative definition of GWB (isotropic, stochastic, Gaussian, unpolarised, and stationary); the true signal produced by the superposition of GWs coming from the ensemble of SBHBs might well be dominated by a handful of signals, therefore significantly departing from isotropy and/or Gaussianity.
The development of detection algorithms targeting multiple individual sources \citep{BabakSesana2012,PetiteauEtAl2013} as well as certain types of anisotropic signals \citep{GairEtAl2014} might prove to be a `more optimal' strategy than searching for a GWB with the aforementioned properties.
\item Predictions should generally be reported in terms of DP and not S/N.
From a frequentist perspective, what matters is the DP at a fixed false alarm probability; a concept which translates into different S/N values for different types of sources and different data analysis techniques.
In our specific case, for fixed DP and FAP, a much larger S/N is required for the detection of a single source than for the detection of a GWB, because of the large number of templates needed in the search of the former.
Moreover, even within the same class of sources, depending on the adopted statistic, fixing DP and FAP may result in a time-dependent S/N threshold (in particular when the GWB becomes no longer a weak signal), which is the case for both our A- and B-statistics{\footnote{We note that this does not apply to the statistic employed by \cite{SiemensEtAl2013}, which is similar to our B-statistic, but with a subtle difference in the definition of the null hypothesis: they assume that all pulsars show a common but uncorrelated red noise (due to some unknown physical process) whose spectral shape mimics exactly that of the GWB. Under this conservative assumption, $\sigma_0=\sigma_1$, and the S/N threshold corresponding to a given DR at a fixed FAP is indeed constant in time. A more detailed study of the statistic used in \cite{SiemensEtAl2013} can be found in \cite{ChamberlinEtAl2015}.}}, as discussed in Section \ref{sc:gwb}.

\item Nonetheless, the S/N behaviour as a function of time encodes interesting information about source detectability.
Both the S/N of a GWB and of a single source increase rapidly with time in the weak signal limit, but after the weak signal regime is abandoned the increase becomes slower.
The transition between the two regimes appears to happen at an S/N $\propto M^{1/2}$, as shown in Figure \ref{fg:snr}. Therefore, adding to a PTA many pulsars with decent precision might be a better strategy to detect {\it any type of signal} than having few ultra-precise pulsars \citep[a point that has also been made by][in the GWB case]{SiemensEtAl2013}, and this issue is certainly worthy of further investigation.
\item The first single sources to be detected will likely be extremely massive ($\mathcal{M}>10^9\msun$), and emitting at a GW frequency of $\approx$10\,nHz.
Incidentally, for typical millisecond pulsar distances of $\sim$kpc, the transition between evolving/non-evolving sources occurs at about that same frequency.
Therefore, it is impossible to discard a priori waveform models and/or detection techniques.
For example, from a frequentist perspective, both ${\cal F}_p$ and ${\cal F}_e$ statistics are equally important tools, and no option should be discarded.
\end{enumerate}

Those results were obtained under a number of simplifying assumption that we now discuss.
\begin{itemize}
\item We assumed circular, GW-driven binaries.
Although this assumption might be reasonable, SBHBs observable with PTAs, are at the centi-to-milliparsec separations, where the coupling to the environment can still play a role.
This fact has been pointed out by a number of authors \citep{KocsisSesana2011,Sesana2013b,McWilliamsEtAl2014,RaviEtAl2014}, and might be relevant for the results reported above.
Efficient environmental coupling drives SBHBs faster towards coalescence; the net effect is that there are less systems contributing at each frequency bin, and the odds that a single source will dominate the signal are likely to increase.
We therefore speculate that environmental coupling, although damaging in terms of GWB detection, might promote the detection of single sources.
On the other hand, environmental coupling might also result in very eccentric binaries \citep{Sesana2010,RoedigSesana2012}.
In this case, each individual system will emit in a whole range of GW harmonics, resulting in a more complicated signal.
The development of data analysis pipelines for eccentric SBHBs has recently begun \citep{ZhuEtAl2015}, but the DP of such systems requires further investigation.
Regardless of the precise mechanism that could affect the actual shape of the GWB, their overall effect could be a lower amplitude of the signal around the lowest frequency bins.
Thus, in practice, subtracting the contribution from the lowest frequency bins (which we did in previous sections in order to simulate the effect of the timing model) could be a sensible way to mimic the reduction in the signal.
In such a case, the DP, plotted in Figure \ref{fig:dp_all}, would be well modelled by the dashed and dotted curves, that disregard the signal from the lowest and two lowest bins, respectively.
\item When evaluating the DP of single sources, we considered only the loudest binary at each frequency bin, and treated the rest of the signal as a stochastic GWB.
However, such approximation might be too crude, since the rest of the signal might be dominated by just a few SBHBs.
\cite{BabakSesana2012} showed that an array of $M$ equally good pulsars can in principle resolve $M/3$ sources at each frequency.
Although the IPTA data set is dominated by a few, very good pulsars, and the resolution of multiple sources seems unlikely, things will be different in the SKA era.
Having a large number of extremely stable pulsars might realistically favour the resolution of multiple SBHBs at each frequency, and, in this respect, the 80\% DP probability for singles sources with an SKA2-type array should be regarded as a robust lower limit.

\item When calculating the detectability of a single source, the sky location of the SBHBs is taken into account; one can see this in Equation (\ref{eq:sssnr_formula}), that depends on the sky coordinates of the binary and the pulsar.
However, the DP of a GWB is insensitive to the particular distribution of the binaries over the sky; the function $S_h$, that encompasses the information about the signal in the DP, depends on the characteristic amplitude of the GWB, made up as the superposition of the GW strains from all SBHBs in the ensemble (summed in quadrature).
This implies that the particular location of the binaries on the celestial sphere is not relevant (only the distance to the binaries is), and the contribution of the ensemble to the cross-correlation is equivalent to that of an isotropic GWB.
In other words, the signal of each binary is isotropically smeared out over the sky.
This can favour the predicted detectability of a GWB over single sources\footnote{We thank Neil Cornish for pointing out this possible caveat.}, since the statistic employed for the detection of the background is optimal for an isotropic signal.
If the ensemble of SBHBs produced a rather anisotropic signal, it could still be detected with a cross-correlation analysis, but the result would be suboptimal \citep{CornishSesana2013}.
In such a case a search for an anisotropic GWB \citep{MingarelliEtAl2013} or a search for multiple sources \citep{BabakSesana2012} would be more efficient, although the performance of an isotropic search or an optimal anisotropic search when the background is anisotropic is likely to be a bit lower than that of an optimal isotropic search for an isotropic background, which could delay detection.
A more sophisticated way to evaluate the detectability of the GWB would require the DP to be sensitive to the sky distribution of binaries, by considering the contribution of each binary to the residuals of each pulsar in the array.
Given the large amount of simulations analysed in this work, such a task would presumably become computationally involved.
We point out, nonetheless, that the DP of the GWB has most of its contribution from the lowest frequency bins, where the background in most models can safely be considered isotropic.
Significant anisotropies, that arise when the background is dominated by fewer binaries, are usually important at rather high frequencies, that do not contribute as much to the DP.
Therefore, we expect that this possible overestimate of the GWB DP should not be significant, and evaluating the detectability of the GWB with a more sophisticated approach would not change the conclusions of this work.

\item We considered ideal millisecond pulsars described by white noise only, ignoring the impact of fitting for a timing model.
This issue was already commented on in previous sections, as well as the possibility of including unmodelled red noise in our estimates.
We defer a more comprehensive treatment of the pulsar noise model to a future work.
\item We considered very idealised PTAs, assuming that the IPTA will simply continue as it is now, without adding any further pulsar, or improving their timing precision.
Moreover, we kept IPTA and SKA separated; a realistic computation of PTA detection probabilities in the SKA era should take into account all previous, valuable IPTA data.
\end{itemize}

Despite all these limitations and caveats, the calculation we presented here is the first quantitative attempt at assessing the `single versus background' issue.
An unresolved GWB is more likely to be detected, but there is a sizeable chance to see a single source first, and the development of the relevant detection pipelines should not be stopped.
An extension of this work to more realistic PTAs and source populations including environmental coupling and eccentricity will be crucial to direct the development of data analysis pipelines and to possibly bring the first GW detection with PTAs closer in time.


\section*{Acknowledgements}
The authors thank the useful help and comments from Bruce Allen, Carsten Aulbert, Stas Babak, Ewan Barr, Neil Cornish, Justin Ellis, Ilya Mandel, Giulio Mazzolo, Antoine Petiteau, Joe Romano, Stephen Taylor, Yan Wang, Karl Wette, and Xingjiang Zhu. AS and JG are supported by the Royal Society through the University Research Fellow scheme. 

\appendix
\section{Optimal statistics for gravitational wave background detection}
\label{app:stats}

\subsection{Cross-correlation}
Here we derive and discuss the properties of the two statistics, A and B (introduced in Section \ref{sc:gwb}), that can be used to characterise the detection of a GWB.
For a filter function of the form $Q(t-t')$, the cross-correlation statistic defined by Equation~(\ref{eq:crosscorr}) can be written in the Fourier domain as
\begin{equation}
S_{ij} = \int _{-\infty}^{\infty} \tilde{Q}(f) \tilde{s}_i^*(f) \tilde{s}_j(f) {\rm d} f,
\end{equation}
which for discretely sampled data becomes a weighted sum over frequency components, $f_k$.
Here the subscripts $i$ and $j$ have been introduced to denote particular pulsars --- $S_{ij}$ is the cross-correlation of the data from pulsar $i$ and pulsar $j$.

The aim is to choose the filter function $Q$ and a suitable combination of the cross-correlations of different pulsar pairs that is `optimal' in the sense that it maximises some quantity that is representative of the probability of detection.
Normally, the optimal filter $Q$ is chosen for a given pair of pulsars first and then these optimal statistics for each pulsar pair are combined to give the final statistic.
However, since the dependence of $S_{ij}$ on $Q$ is linear, if we consider linear combinations of the $S_{ij}$'s these two stages can be done simultaneously and the problem reduces to considering a statistic of the form
\begin{equation}
X = 2 \sum_k \sum_{ij} \lambda_{ijk} s_{ijk},
\label{eq:Xstat}
\end{equation}
and finding the optimal combination of the coefficients $\lambda_{ijk}$.
The factor of $2$ in the preceding equation comes from replacing the integral over both negative and positive frequencies by a sum over positive frequency components only.

The true optimal statistic will be the combination that maximises the DP at a fixed FAP, but as a proxy for this it is usual to consider statistics that maximise the S/N, which is the ratio of the expected value of a statistic in the presence of a signal, $\mu_1$, to its standard deviation.
The standard deviation can either be computed in the absence of a signal, $\sigma_0$, or in the presence of a signal, $\sigma_1$.
The A-statistic was constructed to maximise $\mu_1/\sigma_0$.
This can be thought of as a measure of the significance level at which a given source has a good chance of being detected.
The noise-only standard deviation $\sigma_0$ determines the FAP, and a given FAP therefore corresponds to a certain number of standard deviations, $k \sigma_0$.
Assuming a symmetric distribution, a source with $\mu_1/\sigma_0 = k$ has a $50\%$ chance of being detected if the threshold is set to that FAP.
The B-statistic was constructed to maximise $\mu_1/\sigma_1$.
This is a measure of how inconsistent such a signal is with noise.
The posterior distribution in the presence of a signal will have width $\sigma_1$ and so this statistic measures the number of standard deviations above $0$ the mean of the posterior lies.
As we will see below the two statistics are equivalent in the weak-signal regime, but the B-statistic is more robust for stronger GWBs.

The residuals from each pulsar consist of a signal plus additive noise.
The signal is correlated between different pulsar pairs, but uncorrelated at different frequencies, while the noise is uncorrelated between pulsars and also between frequencies.
In the absence of a signal we therefore have
\begin{eqnarray}
\langle s_{ijk} \rangle_0 &=& 0, \qquad {\rm var}(s_{ijk})_0=P_i(f_k) P_j(f_k) \nonumber \\
{\rm cov}(s_{ijk}, s_{lmn})_0&=&0, \qquad {\rm cov}(s_{ijk}, s_{ilm})_0 = 0
\end{eqnarray}
in which $P_i(f)$ is the power-spectral density of noise in the pulsar $i$.
In the presence of a signal we instead have
\begin{eqnarray}
\langle s_{ijk} \rangle_1 &=& \Gamma_{ij} S_h(f_k), \nonumber \\
{\rm var}(s_{ijk})_1&=&[P_i(f_k)+S_h(f_k)][P_j(f_k)+S_h(f_k)] \nonumber \\
&& + \Gamma_{ij}^2 S_h^2(f_k) \nonumber \\
{\rm cov}(s_{ijk}, s_{lmn})_1&=&[\Gamma_{il} \Gamma_{jm} + \Gamma_{im}\Gamma_{jl}] S_h^2(f_k) \delta_{kn}, \nonumber \\
{\rm cov}(s_{ijk}, s_{ilm})_1 &=& P_i (f_k)\Gamma_{jl} S_h(f_k)  \nonumber \\
&& + [\Gamma_{jl} + \Gamma_{ij}\Gamma_{il}] S_h^2(f_k) \delta_{km}.
\end{eqnarray}
Using these results we can compute $\mu_1$, $\sigma_0$ and $\sigma_1$ for the statistic defined by equation~(\ref{eq:Xstat})
\begin{eqnarray}
\langle X \rangle &=& 2\sum_k \sum_{ij}\lambda_{ijk} \Gamma_{ij} S_h(f_k) \\
{\rm var}(X)_0 &=& 2\sum_k \sum_{ij}\sum_{lm} \lambda_{ijk} [\Sigma_0]_{ij,lm} (f_k) \lambda_{lmk} \\
{\rm var}(X)_1 &=& 2\sum_k \sum_{ij}\sum_{lm} \lambda_{ijk} [\Sigma_1]_{ij,lm} (f_k) \lambda_{lmk}
\end{eqnarray}
in which $\Sigma_0(f_k)$ and $\Sigma_1(f_k)$ are $N_{\rm pp}\times N_{\rm pp}$ covariance matrices for each $f_k$, where $N_{\rm pp}$ is the number of pulsar pairs, labelled by $ij$, and
\begin{eqnarray}
{\Sigma_0}_{ij,ij} (f_k) &=& {\rm var}(s_{ijk})_0, \nonumber \\
 {\Sigma_0}_{ij,lm} (f_k) &=&{\rm cov}(s_{ijk}, s_{lmk})_0
\end{eqnarray}
and similarly for $\Sigma_1$.

\subsection{A-statistic}
Maximising $\mu/ \sigma_0=\langle X \rangle /\sqrt{{\rm var}(X)_0}$ is equivalent to maximising $\langle X \rangle$ subject to the constraint ${\rm var}(X)_0 = 1$, which straightforwardly yields the result
\begin{equation}
\lambda_{ijk} \propto  \sum_{lm} [\Sigma_0^{-1}]_{ij,lm} (f_k) \Gamma_{lm} S_h(f_k).
\end{equation}
The matrices $\Sigma_0(f_k)$ are diagonal and so we obtain the A-statistic
\begin{equation}
X_\text{A}=2 \sum_k \sum_{ij} \frac{\Gamma_{ij} S_{h0}(f_k)} {P_i(f_k) P_j(f_k)} s_{ijk},
\end{equation}
where we have introduced $S_{h0}(f_k)$ to denote the value of the spectral density used to construct the statistic, which is fixed, to distinguish it from $S_h(f_k)$, the actual value of the spectral density in the background, which in general might not be equal to $S_{h0}(f_k)$.
The expected value of the A-statistic in the presence of a signal and its variance in the absence and presence of a signal are then straightforwardly given by
\begin{align}
\label{eq:mu1A}
\mu_1&=2\sum_k \sum_{ij} \frac{\Gamma_{ij}^2 S_h S_{h0}}{P_i P_j}, \\
\label{eq:sigma0A}
\sigma_0^2&=2\sum_k \sum_{ij} \frac{\Gamma_{ij}^2 S_{h0}^2}{P_i  P_j }, \\
\label{eq:sigma1A}
\sigma_1^2&=2\sum_k \sum_{ij} \frac{\Gamma_{ij}^2 S_{h0}^2 \left[[P_i +S_{h}][P_j+S_{h}] +\Gamma_{ij}^2 S_{h}^2 \right]}{\left[P_i P_j \right]^2}.
\end{align}
where it should be assumed that $S_h$, $S_{h0}$ and $P_i$ are evaluated at $f_k$, but we have suppressed the argument for compactness. If $S_{h0}=S_h$ the `A' signal-to-noise ratio, $\mu_1/\sigma_0$, is 
\begin{equation}
\text{S/N}_\text{A}=\left[2 \sum_k \sum_{ij} \frac{\Gamma_{ij}^2 S_h^2}{P_i P_j} \right]^{\frac{1}{2}},
\end{equation}
which is the formula that is most commonly used in the GWB literature.

\subsection{B-statistic}
The equivalent result for the maximisation of $\mu/\sigma_1=\langle X \rangle /\sqrt{{\rm var}(X)_1}$ takes the same form, with $\Sigma_0$ replaced by $\Sigma_1$.
The matrices $\Sigma_1(f_k)$ are not diagonal, but the off-diagonal terms are quadratic in $S_h(f_k)$ and so it is normal to ignore these terms and assume they are sub-dominant relative to the diagonal terms.
This leads to the B-statistic previously given in Equation~(\ref{eq:statBdef}),
\begin{align}
&X_\text{B}= \nonumber \\
&\sum_k \sum_{ij} \frac{2 \Gamma_{ij} S_{h0}(f_k)  s_{ijk}} {[P_i(f_k) +S_{h0}(f_k)][P_j(f_k)+S_{h0}(f_k)] +\Gamma_{ij}^2 S_{h0}^2(f_k)}.
\end{align}
The corresponding expressions for the B-statistic are
\begin{align}
\label{eq:mu1B}
\mu_1&=2\sum_f \sum_{ij} \frac{\Gamma_{ij}^2 S_h S_{h0}}{[P_i+S_{h0}][P_j+S_{h0}]+\Gamma_{ij}^2S_{h0}^2}, \\
\label{eq:sigma0B}
\sigma_0^2&=2\sum_f \sum_{ij} \frac{\Gamma_{ij}^2 S_{h0}^2 P_iP_j}{\left[ [P_i+S_{h0}][P_j+S_{h0}]+\Gamma_{ij}^2S_{h0}^2 \right]^2}, \\
\label{eq:sigma1B}
\sigma_1^2&=2\sum_f \sum_{ij} \frac{\Gamma_{ij}^2 S_{h0}^2 \left[ [P_i+S_{h}][P_j+S_{h}]+\Gamma_{ij}^2S_{h}^2 \right]}{\left[[P_i+S_{h0}][P_j+S_{h0}]+\Gamma_{ij}^2S_{h0}^2 \right]^2},
\end{align}
and for $S_{h0}=S_h$, the `B' signal-to-noise ratio, $\mu_1/\sigma_1$, is
\begin{align}
\label{eq:snrb_bin}
&\text{S/N}_\text{B}=\nonumber\\
&\left[ 2 \sum_f \sum_{ij} \frac{\Gamma_{ij}^2 S_h^2}{P_iP_j+S_h[P_i+P_j]+S_h^2[1+\Gamma_{ij}^2]}\right]^{1/2}.
\end{align}
This formula is not equivalent to Equation (17) of \cite{SiemensEtAl2013} because of the term multiplying $\Gamma_{ij}^2$ in the denominator; numerically, the difference is negligible.
This expression can be derived from Section V.A of \cite{AllenRomano1999}.

\subsection{Comparison}
These two statistics are equivalent at low GWB amplitudes, but can deviate as the amplitude increases.
If we assume that we always adjust the statistic to match the GWB amplitude so that $S_{h0}=S_h$, then $\sigma_0/\sigma_1$ tends to 0 as $S_h \rightarrow \infty$, while $\mu_1/\sigma_1$ tends to a constant value.
From Equation~(\ref{eq:DP}) we see that the DP tends to a finite value less than $1$ as $S_h \rightarrow \infty$.
For the A-statistic the limiting value of $\mu_1/\sigma_1$ is
\begin{equation}
\frac{\mu_1}{\sigma_1} \rightarrow \frac{2\sum_k \sum_{ij} \Gamma_{ij}^2/P_i P_j}{\sqrt{2\sum_k \sum_{ij} \Gamma_{ij}^2[1+\Gamma_{ij}^2]/P_i^2 P_j^2}},
\end{equation}
while for the B-statistic it is
\begin{equation}
\frac{\mu_1}{\sigma_1} \rightarrow \sqrt{2 \sum_k \sum_{ij} \frac{ \Gamma_{ij}^2}{1+\Gamma_{ij}^2}}.
\end{equation}
Both expressions involve sums over all pulsar pairs.
Therefore, if all the $P_i$'s are of similar magnitude, the two expressions scale like $\sqrt{N_{\rm pp}} \sim N_{p}$, and the limiting value of the DP is very close to $1$.
However, if one of the $P_i's$ is much smaller than the others, the A-statistic effectively has fewer terms in it and $\mu_1/\sigma_1$ scales like $\sqrt{N_p}$.
The A-statistic is therefore less robust to having an inhomogeneous set of pulsars and so we use the B-statistic to characterise GWB detection in this analysis.

We finish with three comments.
\begin{enumerate}
\item Firstly, the function $S_{h0}$ appearing in these statistics is not known a-priori.
In practice, this quantity is what we are trying to measure with our observations.
In the case of PTAs, it is most likely that the background will come from a population of merging SBHBs of the type described in this paper and for which the slope should be very close to $S_{h}\propto f^{-13/3}$ and so fixing this as the frequency dependence of $S_{h0}$ is likely to be close to optimal. The amplitude is not known a priori and does not cancel out of the expressions for the B-statistic. One reasonable strategy would be to compute the amplitude of a marginally detectable background, i.e., fix the FAP and compute the background amplitude that yields a $\sim50\%$ probability of detection, and use this in $S_{h0}$. Again this should be near optimal when a background is first detected and will still yield confident detections for louder backgrounds, even though it will be sub-optimal at such amplitudes. Other approaches, such as using a set of statistics constructed with a range of different functions $S_{h0}$, or an iterative approach in which an estimate of the amplitude obtained from the measured data is used to contract a new statistic and so on, would also be possible. An investigation of these approaches to data analysis is beyond the scope of this paper. However, we found that, in practice, the performance of the statistic did not depend strongly on the assumption about $S_{h0}$ and so for all the calculations in this paper we assume perfect knowledge of the background and set $S_{h0}=S_h$.
\item  Secondly, in deriving the B-statistic we ignored the off-diagonal elements of the cross-correlation matrix.
The B-statistic will be valid in an intermediate signal regime, but will become sub-optimal at large background amplitudes due to the omission of these terms.
This was also discussed in~\cite{AllenRomano1999}, where it was argued that these terms were always subdominant for observations with GBDs.
The importance of these terms has not been investigated for PTAs.
\item Finally, neither the A-statistic nor the B-statistic is truly optimal in the sense of maximising the DP for a fixed FAP.
Not only have off-diagonal correlations been ignored in deriving the B-statistic, but the procedure of maximising $\mu_1/\sigma_0$ or $\mu_1/\sigma_1$ is not equivalent to maximising the DP.
Under a Gaussian assumption we need to maximise $[\mu_1 - k\sigma_0]/\sigma_1$ where $k$ is a constant that depends on the chosen FAP.
This maximisation can be done, although it is more difficult, and we did not consider it here.
Simulations indicate that both the A- and B-statistics are nearly optimal in the weak signal limit, while the B-statistic continues to be near optimal for higher background amplitudes.
\end{enumerate}

\bibliographystyle{mn2e}
\bibliography{snr_computation}

\end{document}